\documentclass[preprint,nofootinbib,aps,superscriptaddress,eqsecnum]{revtex4-2} 
\usepackage[colorlinks,citecolor=purple]{hyperref}
\usepackage[usenames]{color}
\usepackage{xcolor}
\usepackage{float}
\pdfoutput=1
\usepackage{extarrows}
\usepackage{tabularx}
\usepackage{graphicx}
\usepackage{amsmath}
\usepackage{amsfonts}
\usepackage{amssymb}
\usepackage[justification=centering]{caption} 
\usepackage{caption}
\usepackage{subcaption}
\usepackage{color}
\allowdisplaybreaks
\def\bea{\begin{eqnarray}}
	\def\eea{\end{eqnarray}}
\def\be{\begin{equation}}
	\def\ee{\end{equation}}

\begin{document}
	\title{$M_{W_R}$ dependence of leptogenesis in Minimal Left-Right Symmetric Model with different strengths of Type-II seesaw mass.}
	\author{Ankita Kakoti}
	\email{ankitak@tezu.ernet.in}
	\author{Mrinal Kumar Das}
	\email{mkdas@tezu.ernet.in}
	\affiliation{Department of Physics, Tezpur University, Tezpur 784028, India}
	\begin{abstract}
		Left Right Symmetric Model (LRSM) being an extension of the Standard model of particle physics incorporates within itself Type-I and Type-II seesaw mass terms naturally. Both the mass terms can have significant amount of contribution to the resulting light neutrino mass within the model and hence on the different phenomenology associated within. In this paper, we have thoroughly analyzed and discussed the implications of specifying different weightages to both the mass terms and also the study has been carried out for different values of $M_{W_R}$ which is mass of the right-handed gauge boson. This paper also gives a deeper insight into the new physics contributions of Neutrinoless Double Beta Decay $(0\nu\beta\beta)$ and their variations with the net baryon asymmetry arising out of the model. Therefore, the main objective of the present paper rests on investigating the implications of imposing different weightage to the type-I and type-II seesaw terms and different values of $M_{W_R}$ on the new physics contributions of $0\nu\beta\beta$ and net baryon asymmetry arising out as a result of resonant leptogenesis. LRSM in this work has been realized using modular group of level 3, $\Gamma(3)$ which is isomorphic to non-abelian discrete symmetry group $A_4$, the advantage being the non-requirement of flavons within the model and hence maintaining the minimality of the model.
	\end{abstract}
\maketitle
\newpage
\begin{center}
	\section{\label{lrsm1}\textbf{Introduction}}
\end{center}

Considering the fact that neutrinos are massive and mixing occur between their flavors which has been proved by the neutrino oscillation experiments, we are compelled to move beyond the Standard paradigm. Although Standard Model (SM) has been enormously victorious in explaining the fundamental particles and their interactions, it fails to address some issues like the origin of tiny neutrino masses, baryon asymmetry of the universe (BAU) \cite{Kolb:1979ui}, lepton flavor violation (LFV) \cite{Altarelli:2012bn}, origin of dark matter etc \cite{Taoso:2007qk,Turner:1993dra}.\\In this context, we can state that the seesaw mechanism \cite{Minkowski:1977sc,Schechter:1980gr,Mohapatra:1979ia,Wetterich:1981bx,Antusch:2004xy,Foot:1988aq,Haba:2016lxc} has proven to be a very straightforward way to explain the tiny neutrino masses, the two main componets of the mechanism being, addition of right-handed (RH) neutrinos to the SM and endowing the RH neutrinos with a Majorana mass which breaks the accidental global (B-L) symmetry of the SM. However, in the context of electroweak gauge theory of the SM described by $SU(3)_C \otimes SU(2)_L \otimes U(1)_Y$, there is no underlying principle that explains this mechanism, but rather they are put in by hand. In this context, we have two ultraviolet-complete theories where both the components of seesaw arise in a natural manner and these are,
\begin{itemize}
	\item The left-right (LR) symmetric theories of weak interactions based on the gauge group $SU(3)_C \otimes SU(2)_L \otimes SU(2)_R \otimes U(1)_{B-L}$ \cite{Mohapatra:1979ia,Senjanovic:1978ev,Senjanovic:2018xtu,BhupalDev:2018xya,Grimus:1993fx,Mohapatra:1974gc,Pati:1974yy,Senjanovic:1975rk}.
	\item  $SO(10)$ grand-unified theory for all interations \cite{Pernow:2019tuf}.
\end{itemize}
In this work, our focus is on low-scale Left-Right symmetric model (LRSM) where the seesaw scale can be in a few TeV range, such that it is accessible to the LHC, while satisfying the observed charge-lepton and neutrino mass spectra.\\
One important question that arises here is that why the seesaw mechanism is so favourable for the explanation of tiny neutrino masses. This is because, when we take seesaw mechanism into consideration, the same Yukawa that are responsible for generating small neutrino masses may also be helpful in  resolving one of the outstanding puzzles of cosmology, namely, the origin of matter-antimatter asymmetry, via leptogenesis. Leptogenesis can be picturized by the out-of equilibrium decays of the RH neutrinos via the modes $N\rightarrow L_{l}\phi$ and $N\rightarrow L_{l}^c\phi^c$, where, $L_l = (\nu_L l)_{L}^T$ are the $SU(2)_L$ lepton doublets, $\phi$ is the Higgs doublet and the superscript c denotes the CP conjugate. In the presence of CP violation in the Yukawa sector, the decays mentioned above can lead to lepton asymmetry in the early universe satisfying the three Sakharov conditions. This asymmetry then undergoes thermodynamic evolution with the expansion of the universe and different reactions present in the model have their impact on washing out part of the asymmetry. The remaining final lepton asymmetry is converted to baryon asymmetry via sphaleron processes before the electroweak phase transition.\\
For TeV scale seesaw models, we require resonant leptogenesis (RL) for the generation of adequate lepton asymmetry, where atleast two of the heavy Majorana neutrinos have a small difference comparable to their decay widths. The TeV scale new particles in LRSM also leads to interesting collider signatures.\\
In this work, we have also taken into account the study of neutrinoless double beta decay (NDBD)  \cite{Vergados:2012xy}, \cite{Cardani:2018lje}, because the possible observation of NDBD would play an important role in understanding the origin of BAU as NDBD would imply that lepton number is indeed not conserved, which is one of the essential conditions for leptogenesis. In LRSM, there are several contributions to NDBD that involve left-right sectors individually and also some contributions that involves both the sector through left-right mixing accompanied by both light and heavy neutrinos. Left-Right mixing is a ratio of the Dirac and Majorana mass scales $[M_{D}M_{RR}^{-1}]$, appearing in the type-I seesaw mass formula as we will discuss later. While taking to account large left-right mixing we have to focus on the momentum dependent contributions of NDBD, that is, the contribution that arises as a result of the mixed diagrams with simultaneous mediation of $W_L$ and $W_R$ accompanied by light left-handed and heavy right-handed neutrino known as $\lambda$ and $\eta$ contributions.\\
In this work, we have firstly realized LRSM with the help of modular $A_4$ symmetry and thus obtained the mass matrices in terms of the modular forms $(Y_1,Y_2,Y_3)$, and then we move forward to study the linkage between the different contributions of NDBD and baryogenesis. We have considered different strengths of the type-I and type-II seesaw masses with differents values of the $SU(2)_R$ breaking scale, namely, type-II seesaw mass being $30\%,50\%,70\%$ dominant for $M_{W_R} = 1.5,2.2,2.5,3,5,10,18$ TeV. We have not considered any particular texture for the Dirac and Majorana mass matrices and in order to account for the neutrino mass of the order of sub eV, the Dirac mass that we have determined is in the scale of MeV, as the heavy right handed neutrino is kept in the TeV scale. This leads to not so large left-right mixing and hence we get non-negligible effects of the momentum dependent contributions of NDBD. Although we have studied all the contributions of NDBD, however to relate baryogenesis to the phenomena, we have considered only the contributions appearing from the mediation of heavy right-handed neutrino and the momentum dependent $\lambda$ and $\eta$ contributions. Since the effective mass governing NDBD is dependent on the Majorana phases, $\alpha$ and $\beta$, it would be compelling to study if there is a link between NDBD and BAU. So, this paper has been organized by firstly discussing about LRSM and the origin of neutrino mass in section \ref{lrsm2}, section \ref{lrsm3} describing the realization of LRSM with $A_4$ modular symmetry and section \ref{lrsm4} describes the phenomenon of resonant leptogenesis and NDBD within LRSM, section \ref{lrsm5} discusses the numerical analysis and results and we conclude our work by discussion and conclusion in section \ref{lrsm6} and \ref{lrsm7}. The appendices discussed the minimisation of the Higgs potential for the model and also introduces the concept of modular symmetry which has been used for the realisation of the model.
\begin{center}
	\section{Left-Right Symmetric Model and Neutrino mass.}\label{lrsm2}
\end{center}

 Left-Right Symmetric Model (LRSM) is described by the gauge group $SU(3)_C \otimes SU(2)_L \otimes SU(2)_R \otimes U(1)_{B-L}$. As is evident from the gauge group that unlike SM, both the left-handed and right handed particles are taken as doublets in the said model. The scalar sector of the model consists of one Higgs bidoublet $\phi (1,2,2,0)$ and two scalar triplets given by $\Delta_L(1,3,1,2)$ and $\Delta_R(1,1,3,2)$. Because of the presence of right-handed (RH) neutrinos and scalar triplets, LRSM incorporates within itself $Type-I$ and $Type-II$ seesaw masses naturally.\\
The fermions can attain mass only when the necessary Yukawa Lagrangian is constructed which shows coupling of particles to the bidoublet $\phi$. The Yukawa Lagrangian for fermions incorporating the bidoublet is given by,
\begin{equation}
	\label{E:1}
	\mathcal{L_{D}} = \overline{l_{iL}}(Y_{ij}^l \phi + \widetilde{Y_{ij}^l}\widetilde{\phi})l_{jR}+ h.c   
\end{equation}
where, $l_L$ and $l_R$ are the left-handed and right-handed lepton fields. $Y^l$ being the Yukawa coupling corresponding to leptons and $Y^q$ being the Yukawa coupling for the quarks. 
The Yukawa Lagrangian where the scalar triplets play a role in providing Majorana mass to the neutrinos is given by,
\begin{equation}
	\label{E:2}
	\mathcal{L_M}=f_{L,ij}{\Psi_{L,i}}^TCi\sigma_2\Delta_L\Psi_{L,j}+f_{R,ij}{\Psi_{R,i}}^TCi\sigma_2\Delta_R\Psi_{R,j}+h.c
\end{equation}
$f_L$ and $f_R$ are the Majorana Yukawa couplings and are equal because of the discrete left-right symmetry. The family indices $i,j$ runs from $1$ to $3$ representing the three generations of the fermions. $C=i\gamma_{2}\gamma_{0}$ is the charge conjugation operator, where $\gamma_{\mu}$ are the Dirac matrices and  $\tilde{\phi} = \tau_{2}\phi^{*}\tau_{2}$.\\The symmetry breaking of the gauge group describing the model takes place in two steps. The model gauge group is first broken down to the Standard Model gauge group by the VEV of the scalar triplet $\Delta_R$, and then the Standard Model gauge group is broken down to the electromagnetic gauge group i.e., $U(1)_{em}$ by the VEV of the bidoublet and a tiny vev of the scalar triplet $\Delta_L$.\\So the  resultant light neutrino mass of LRSM is expressed as a sum of the type-I and type-II seesaw mass terms, given as,
\begin{equation}
	\label{E:3}
	M_{\nu} = M_{\nu}^I + M_{\nu}^{II}
\end{equation}
where,
\begin{equation}
	\label{E:4}
	M_{\nu}^I = M_{D}M_{R}^{-1}M_{D}^T
\end{equation}
is the type-I seesaw mass, and type-II seesaw mass is given by
\begin{equation}
	\label{E:5}
	M_{\nu}^{II} = M_{LL}
\end{equation}
$M_D$ is the Dirac mass matrix and $M_R$ is the right-handed Majorana mass matrix, where, $M_R = \sqrt{2}v_{R}f_{R}$ and $M_L = \sqrt{2}v_{L}f_{L}$. $v_R$ and $v_{L}$ are the respective VEVs of $\Delta_R$ and $\Delta_L$. The magnitudes of the VEVs follows the relation, $|v_L|^2 < |k^{2} +k'^{2}| < |v_R|^2$.  In LRSM however, the type-I and type-II mass terms can be expressed in terms of the heavy right-handed Majorana mass matrix, so Eq.\eqref{E:3} will follow,
\begin{equation}
	\label{E:6}
	M_\nu = M_D M_{R}^{-1} M_D^T + \gamma\Biggl(\frac{M_W}{v_R}\Biggl)^2 M_{RR}
\end{equation}
where, $\gamma$ is a dimensionless parameter which is a function of various couplings, appearing in the VEV of the triplet Higgs $\Delta_L$, i.e., $v_L = \gamma (\frac{v^2}{v_R})$ and here, $v = \sqrt{k^2 + k'^2}$, and
\begin{equation}
	\label{E:7}
	\gamma =  \frac{\beta_1 k k' + \beta_2 k^2 + \beta_3 k'^2}{(2\rho_1 - \rho_3)(k^2 + k'^2)}
\end{equation} 
In our model, the dimensionless parameter $\gamma$ has been fine tuned to $\gamma \approx 10^{-13}$ and $v_R$ is of the order of $TeV$. $\beta_i$ and $\rho_i$ are dimensionless parameters which appear in the potential of the model which has been thoroughly elaborated in the appendix.\\
The light neutrino mass obtained can be expressed in terms of a matrix given as,
\begin{equation}
	M_{\nu}=\begin{pmatrix}
		M_{LL} & M_{D}\\
		M_{D}^{T} & M_{RR}
	\end{pmatrix}
\end{equation}
This matrix is a $6 \times 6$ matrix which can be diagonalized by a unitary matrix as follows,
\begin{equation}
	\label{E:22}
	\nu^{T}M_{\nu}\nu = \begin{pmatrix}
		\hat M_{\nu} & 0\\
		0 & \hat M_{RR}
	\end{pmatrix}
\end{equation}
where, $\nu$ represents the diagonalizing matrix of the full neutrino mass matrix, $M_{\nu}$,$\hat{M_{\nu}} = diag(m_1,m_2,m_3)$, with $m_i$ being the light neutrino masses and $\hat{M_{RR}} = diag(M_1,M_2,M_3)$, with $M_i$ being the heavy right-handed neutrino masses.\\
The diagonalizing matrix can be represented as,\\
\begin{equation}
	\label{E:23}
	\nu = \begin{pmatrix}
		U & S\\
		T & V
	\end{pmatrix} \approx \begin{pmatrix}
	1-\frac{1}{2}RR^\dagger & R\\
	-R^\dagger & 1-\frac{1}{2}R^\dagger R
\end{pmatrix} \begin{pmatrix}
V_{\nu} & 0\\
0 & V_R
\end{pmatrix}
\end{equation}
where, $R$ describes the left-right mixing and is given by,\\
\begin{equation}
	\label{E:24}
	R = M_{D}M_{RR}^{-1} + O(M_{D}^3(M_{RR}^{-1})).
\end{equation}
The matrices $U,V,S$ and $T$ are as follows,
\begin{equation}
	\label{E:25}
	U = [1-\frac{1}{2}M_{D}M_{RR}^{-1}(M_D M_{RR}^{-1})^\dagger]V_{\nu}
\end{equation}
\begin{equation}
	\label{E:26}
		V = [1-\frac{1}{2}(M_{D}M_{RR}^{-1})^{\dagger} M_D M_{RR}^{-1}]V_{\nu}
\end{equation}
\begin{equation}
	\label{E:27}
		S = M_D M_{RR}^{-1} v_{R} f_{R}
\end{equation}
\begin{equation}
	\label{E:28}
		T = -(M_D M_{RR}^{-1})^\dagger V_{\nu}
\end{equation}

In this work, realization of LRSM has been done with the help of $A_4$ modular symmetry, which has been discussed thoroughly in the next section.
\newpage
 \section{\label{lrsm3}Realization of LRSM with modular group $\Gamma(3)$ having weight 2 .}

 It has been realized since a long course of time that symmetries do hold an important position in the context of particle physics. LRSM has been constructed with the help discrete flavor symmetries in several earlier works like \cite{Duka:1999uc}, \cite{Boruah:2022bvf},\cite{Sahu:2020tqe},\cite{Rodejohann:2015hka}. In our present work, we have realized the model using $A_4$ modular symmetry (non-SUSY case). The advantage of using modular symmetry rather than flavor symmetry is that we do not require the use of any flavons within the model and hence it helps in keeping the model minimal. The model contains usual particle content of LRSM \cite{Sahu:2020tqe}. The lepton doublets transform as triplets under $A_4$ and the bidoublet and scalar triplets transform as 1 under $A_4$ \cite{Abbas:2020qzc}. As we have considered modular symmetry, we assign modular weights to the particles, keeping in mind that matter multiplets corresponding to the model can have negative modular weights, but the modular forms cannot be assigned negative weights. We assign modular weights to the particles in a manner that the sum of the modular weights of the particles in each term of the Lagrangian sum up to be zero. Modular weights corresponding to each particle is shown in table \ref{tab:Table 1}. The Yukawa Lagrangian for the leptonic sector in LRSM is given by Eq.\eqref{E:1},\eqref{E:2} from which we can write the Yukawa Lagrangian of our $A_4$ modular symmetric LRSM, for the lepton sector, by introducing Yukawa coupling in the form of modular forms $Y$ is given as,
\begin{equation}
	\label{e:25}
	\mathcal{L_Y} = \overline{l_L}\phi{l_R}Y+\overline{l_L}\tilde{\phi}{l_R}Y+{{l_R}^T}C i{\tau_2}{\Delta_R}{l_R}{Y}+{{l_L}^T}C i{\tau_2}{\Delta_L}{l_L}{Y}
\end{equation}
\begin{table}[H]
	\begin{center}
		\begin{tabular}{|c|c|c|c|c|c|c|}
			\hline
			Gauge group & $l_L$ & $l_R$ & $\phi$ & $\Delta_L$ & $\Delta_R$ \\
			\hline
			$SU(3)_C$ & 1 & 1 & 1 & 1 & 1\\
			\hline
			$SU(2)_L$ & 2 & 1 & 2 & 3 & 1 \\
			\hline
			$SU(2)_R$ & 1 & 2 & 2 & 1 & 3 \\
			\hline
			$U(1)_{B-L}$ & -1 & -1 & 0 & 2 & 2 \\
			\hline
			$A_4$ & 3 & 3 & 1 & 1 & 1 \\
			\hline 
			$k_I$ & 0 & -2 & 0 & -2 & 2 \\
			\hline
		\end{tabular}
		\caption{\label{tab:Table 1}Charge assignments for the particle content of the model.}
	\end{center}
\end{table}

The Yukawa couplings $Y = (Y_1,Y_2,Y_3)$ are expressed as modular forms of level 3.
\begin{table}[H]
	\begin{center}
		\begin{tabular}{|c|c|}
			\hline
			& Y (modular forms)\\
			\hline
			$A_4$ & $3$ \\
			\hline
			$k_I$ & $2$ \\
			\hline
		\end{tabular}
		\caption{\label{tab:Table 2}Charge assignment and modular weight for the corresponding modular Yukawa form for the model.}
	\end{center}
\end{table}
Table \ref{tab:Table 2} shows the charge assignment for the modular form of level 3 and its corresponding modular weight.
In our work, we are concerned with the mass of the neutrinos and as such, using $A_4$ modular symmetry and using the multiplication rules for $A_4$ group \cite{Ding:2013bpa}, we construct mass matrices as given below.\\
The Dirac mass matrix from the Yukawa Lagrangian has been determined in terms of $(Y_1,Y_2,Y_3)$ as,
\begin{equation}
	\label{e:a1}
	M_D =v\begin{pmatrix}
	2Y_1 & -Y_3 & -Y_2\\
	-Y_2 & -Y_1 & 2Y_3\\
	-Y_3 & 2Y_2 & -Y_1
\end{pmatrix}
\end{equation}
where, $v$ is the VEV of the Higgs bidoublet $\phi$.\\It is to be noted here that the textures of the Dirac mass matrix and also the right-handed mass matrix has been determined, so it is a fact that the type-I seesaw mass texture can be determined. But, the pivotal investigation in the present study revolves around the assignment of different strengths to the type-II seesaw mass term and hence from there, the value of the type-I seesaw mass has been calculated in order to determine the light neutrino mass.\\
The right-handed Majorana mass matrix is given by,
\begin{equation}
	\label{e:a}
	M_R =v_R\begin{pmatrix}
		2Y_1 & -Y_3 & -Y_2\\
		-Y_3 & 2Y_2 & -Y_1\\
		-Y_2 & -Y_1 & 2Y_3
	\end{pmatrix}
\end{equation}
where, $v_R$ is the VEV for the scalar triplet $\Delta_R$. As it is seen that the Majorana mass matrix for our model is found to be symmetric in nature as it should be. Under these assumptions for modular symmetric LRSM and the basis that we have considered, our charged lepton mass matrix is also found to be diagonal.\\
The type-II seesaw mass is given by,
\begin{equation}
	\label{e:c}
	M_{\nu}^{II}=M_{LL}
\end{equation}
where, 
\begin{equation}
	\label{e:d}
	M_{\nu}^{II} = v_L\begin{pmatrix}
		2Y_1 & -Y_3 & -Y_2\\
		-Y_3 & 2Y_2 & -Y_1\\
		-Y_2 & -Y_1 & 2Y_3
	\end{pmatrix}
\end{equation}
where,$v_L$ is the vev of the scalar triplet $\Delta_L$.\\
The determination of the values of the modular forms $(Y_1,Y_2,Y_3)$ has been done using the relations as mentioned in the section of numerical analysis and results.\\

\section{\label{lrsm4}Resonant leptogenesis and neutrinoless double beta decay $(0\nu\beta\beta)$ in TeV scale LRSM.}

\begin{figure}[H]
	\centering
	\includegraphics[scale=0.4]{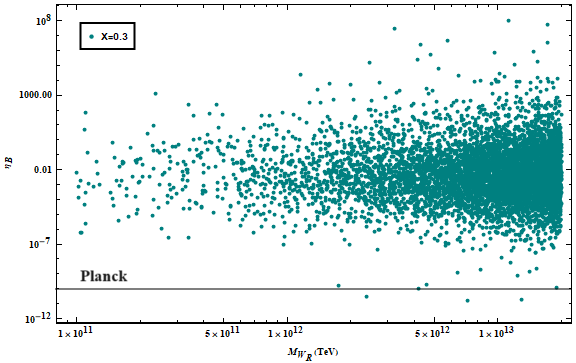}
	\includegraphics[scale=0.4]{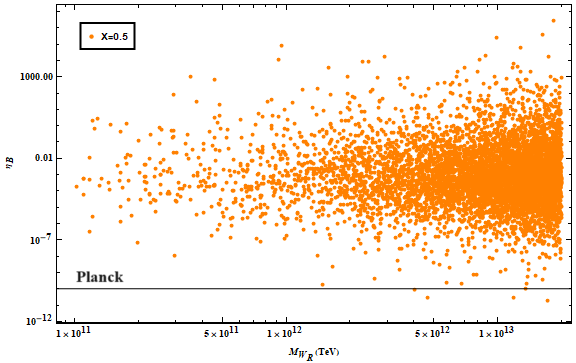}
	\includegraphics[scale=0.4]{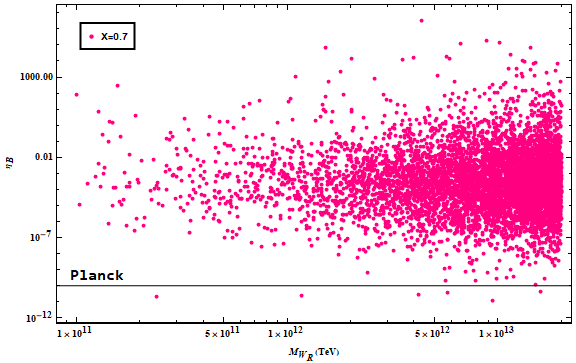}
	\caption{\label{f:1a}Variation of $\eta_{B}$ with $M_{W_R}$ for normal hierarchy(NH) for $X=0.3,0.5,0.7$. The black line indicates the Planck limit on $\eta_B$.}
\end{figure}

\begin{figure}[H]
	\centering
	\includegraphics[scale=0.4]{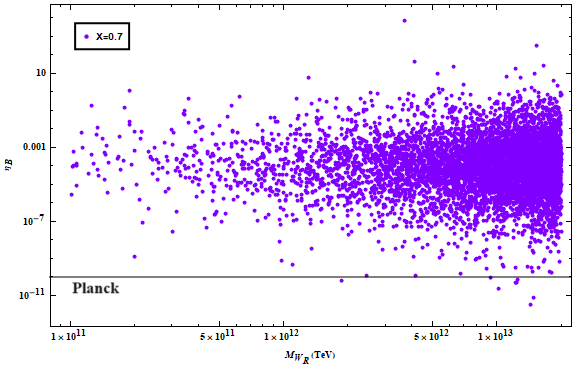}
	\caption{\label{f:1b}Variation of $\eta_{B}$ with $M_{W_R}$ for inverted hierarchy(IH) for $X=0.3,0.5,0.7$. The black line indicates the Planck limit on $\eta_B$.}
\end{figure}

In TeV scale seesaw models, an adequate amount of lepton asymmetry can be generated by the phenomenon of resonant leptogenesis (RL) \cite{Asaka:2018hyk,Blanchet:2009bu,Flanz:1996fb,Dev:2015cxa}. For RL, two of the right-handed Majorana neutrinos need to be degenerate which has been found to be so in the present work. The presence of RH neutrinos and scalar triplets in the framework of LRSM suggests that their decays can give rise to lepton asymmetry. The net lepton asymmetry will therefore be generated due to the two seesaw terms. So, in this work, we investigate the contribution of both the seesaw terms towards the lepton asymmetry, assigning them different weightages (e.g. 30\%,50\% and 70\%).\\For RL however there is a basic requirement that the a pair of Majorana neutrinos must have a mass difference comparable to their decay widths which will be illustrated later in the work.\\ The rate at which the RH neutrinos decay is governed by the Yukawa couplings, and is hence given by \cite{Borgohain:2017akh},
\begin{equation}
	\label{E:8}
	\Gamma_{i}=(Y_{\nu}^{\dagger}Y_{\nu})_{ii} \frac{M_i}{8\pi}
\end{equation}
As already mentioned previously that one important condition for RL is that the mass difference of the two heavy RH neutrinos must be comparable to their decay width, i.e., $M_{i}-M_{j}=\Gamma$. In such a case, the CP asymmetry may become very large. The CP violating asymmetry is thus given by \cite{Xing:2015fdg},
\begin{equation}
	\label{E:9}
	\epsilon_{i}=\frac{Im[(Y_{\nu}^{\dagger}Y_{\nu})_{ij}^2]}{(Y_{\nu}^{\dagger}Y_{\nu})_{11}(Y_{\nu}^{\dagger}Y_{\nu})_{22}}.\frac{(M_{i}^{2}-M_{j}^{2})M_{i}\Gamma_{j}}{(M_{i}^{2}-M_{j}^{2})^{2}+M_{i}^{2}\Gamma_{j}^{2}}
\end{equation}
The variables $i,j$ run over 1 and 2 and $i\neq j$.\\
The CP asymmetries $\epsilon_{1}$ and $\epsilon_{2}$ can give rise to a net lepton asymmetry, provided the expansion rate of the universe is larger than $\Gamma_{1}$ and $\Gamma_{2}$. This can further be converted into baryon asymmetry of the universe by $B+L$ violating sphaleron processes.\\
Now, due to the presence of several new heavy particles in LRSM, along with the standard light neutrino contribution to $0\nu\beta\beta$, several new physics contributions also come into the picture. There are total eight contributions of the phenomenon to LRSM. They are mentioned as under,
\begin{itemize}
	\item Standard Model contribution to NDBD where the intermediate particles are the $W_L$ bosons and light neutrinos, the process in which the amplitude depends upon the leptonic mixing matrix elements and light neutrino masses.
	\item Heavy right-handed neutrino contribution in which the mediator particles are the $W_L$ bosons and the amplitude depends upon the mixing between light and heavy neutrinos as well as the mass of the heavy neutrino.
	\item  Light neutrino contribution to NDBD where the intermediate particles are $W_R$ bosons and the amplitude depends upon the mixing between light and heavy neutrinos as well as mass of the right-handed gauge boson $W_R$.
	\item  Heavy right-handed neutrino contribution where the mediator particles are the $W_R$ bosons. The amplitude of this process is dependent on the elements of the right handed leptonic mixing matrix and mass of the right-handed gauge boson, $W_R$ as well as the mass of the heavy right handed Majorana neutrino.
	\item  Light neutrino contribution from the Feynman diagram mediated by both $W_L$ and $W_R$, and the amplitude of the process depends upon the mixing between light and heavy neutrinos, leptonic mixing matrix elements, light neutrino masses and the mass of the gauge bosons, $W_L$ and $W_R$ ($\lambda$ contribution). 
	\item Heavy neutrino contribution from the Feynman diagram mediated by both $W_L$ and $W_R$, and the amplitude of the process depends upon the right handed leptonic mixing matrix elements, mixing between the light and heavy neutrinos, also the mass of the gauge bosons, $W_L$ and $W_R$ and the mass of the heavy right handed neutrino ($\eta$ contribution).
	\item Scalar triplet contribution ($\Delta_L$) in which the mediator particles are $W_L$ bosons, and the amplitude for the process depends upon the masses of the $W_L$ bosons, left-handed triplet Higgs, as well as their coupling to leptons.
	\item Right-handed scalar triplet contribution ($\Delta_R$) contribution to NDBD in which the mediator particles are $W_R$ bosons, and the amplitude for the process depends upon the masses of the $W_R$ bosons, right-handed triplet Higgs, $\Delta_R$ as well as their coupling to leptons. 
\end{itemize}
In this work, the momentum dependent, that is, $\lambda$ and $\eta$ contributions of $0\nu\beta\beta$ have been considered. The standard light neutrino contribution and new physics contributions by the heavy RH neutrino and scalar triplet contribution in modular symmetric LRSM have already been done in one of our previous work \cite{Kakoti:2023isn}.
\begin{center}
	\section{\label{lrsm5}Numerical Analysis and Results}
\end{center}
The light neutrino mass in LRSM is given by,
\begin{equation}
	\label{E:10}
	M_\nu = M_\nu^{I} + M_\nu^{II}
\end{equation}
where, the type-I seesaw mass term is,
\begin{equation}
	\label{E:11}
	M_\nu^{I} = M_{D}M_{R}^{-1}M_{D}^{T}
\end{equation}
and, \begin{equation}
	\label{E:12}
	M_\nu^{II} = M_{LL}
\end{equation}
In this work, we aim to study the strengths of each seesaw terms that contribute towards the light neutrino mass and hence the associated phenomenology. So, we have considered,
\begin{equation}
	\label{E:13}
	M_\nu^{II(diag)} = XM_\nu^{(diag)}
\end{equation}
where, the parameter X decides the relative strength of the type-II seesaw mass term towards the resulting light neutrino mass. For example, considering that the type-II mass term contributes $10\%$ towards the light neutrino mass, then,
\begin{equation}
	X = 10\% = \frac{10}{100} = 0.1
\end{equation} 
In this work, we have considered the contributions of type-II seesaw mass terms towards the light neutrino mass as, $X=0.3,0.5,0.7$ that is, contribution from type-II seesaw term is considered to be $30\%,50\%,70\%$ respectively. \\
As we are using modular symmetry in LRSM, so our mass matrices are expressed in terms of the modular forms $(Y_1,Y_2,Y_3)$. So, the type-II mass matrix is expressed as,
\begin{equation}
	\label{E:14}
M_\nu^{II}=v_L \begin{pmatrix}
	2Y_1 & -Y_3 & -Y_2 \\
	-Y_3 & 2Y_2 & -Y_1 \\
	-Y_2 & -Y_1 & 2Y_3
\end{pmatrix}
\end{equation}
where, $v_L$ is of the order of $eV$.\\
Using \eqref{E:13}, we can determine the values of the modular forms and hence further we can calculate the values of the neutrino masses and take into consideration effective masses of different contributions of $0\nu\beta\beta$ in the present work. In LRSM, from type-II seesaw term, $M_{R}$ can be written as,
\begin{equation}
	\label{E:15}
	M_{R}=\frac{1}{\gamma}\Biggl(\frac{v_R}{M_{W_{L}}}\Biggl)^{2}M_\nu^{II}
\end{equation}
As already mentioned that $\gamma$ in this work has been fine tuned to $10^{-13}$. We have considered different values of the $SU(2)_R$ breaking scale $v_R$ for our further analysis, namely $1.5,2.2,2.5,3,5,10$ and $18$ TeV. These values will be useful for us to determine the common parameter space of the phenomena being considered like, baryon asymmetry of universe (BAU), contributions of $0\nu\beta\beta$ etc. The mass of the left-handed gauge boson is taken as, $M_{W_{L}} = 80GeV$.\\
To proceed further into our study and for the required analysis, firstly we need to determine the values of Yukawa couplings expressed as modular forms $(Y_1,Y_2,Y_3)$. In order to determine their values, we make use of Eq.\eqref{E:13}. For that, we first check whether the type-II mass matrix $M_{\nu}^{II}$ is diagonalizable. In the present work, $M_{\nu}^{II}$ is indeed diagonalizable and so we diagonalize the matrix to obtain $M_{\nu}^{II(diag)}$ in terms of $(Y_1,Y_2,Y_3)$. The RHS (right-hand side) of Eq.\eqref{E:13} contains the parameters whose values are known, that is $X$ is taken to be $0.3,0.5$ and $0.7$, also $M_{\nu}^{diag}$ = $diag (m_1,m_2,m_3)$ is the diagonal light neutrino mass matrix. Equating each of the elements corresponding to the derived matrices on the RHS and LHS of Eq.\eqref{E:13} has helped us obtain the values of  $(Y_1,Y_2,Y_3)$. 
Now, we substitute these values into the obtained mass matrices so as to determine the required parameters associated with the model. The range of the Yukawa couplings for both normal and inverted hierarchies have been found to be in the order of $O(10^{-4})$ to $O(1)$.
We then determine the values of the leptonic CP asymmetry $\epsilon_{1}$ and $\epsilon_{2}$, by using \eqref{E:9}, where $Y_v = \frac{M_D}{v}$, $M_D$ being the Dirac mass and $v$ is the VEV of the Higgs bidoublet and is in $GeV$. The decay rates in \eqref{E:9} can be calculated using the relation \eqref{E:8}.\\
The CP violating asymmetries $\epsilon_{1}$ and $\epsilon_{2}$ can give rise to net lepton number asymmetry, provided the expansion rate of the universe is larger than $\Gamma_{1}$ and $\Gamma_{2}$.The net baryon asymmetry is then calculated using the following relation \cite{Buchmuller:2004tu},
\begin{equation}
	\label{E:16}
	\eta_B \approx -0.96 \times 10^{-2}\sum_{i}(k_{i}\epsilon_{i})
\end{equation}
$k_i$ being the efficiency factors measuring the washout effects. Some parameters are needed to defined as,
\begin{equation}
	\label{E:17}
	K_i \equiv \frac{\Gamma_{i}}{H}
\end{equation}
Equation \eqref{E:17} is defined at temperature $T=M_{i}$. The Hubble's constant is given by, $H\equiv \frac{1.66\sqrt{g_{*}}T^{2}}{M_{Planck}}$, where, $g_{*}=107$ and $M_{Planck} = 1.2 \times 10^{19}$ GeV is the Planck mass. decay width is estimated using \eqref{E:8}. The efficiency factors $k_i$ can be calculated using the formula \cite{Blanchet:2008pw},
\begin{equation}
	\label{E:18}
	k_1 \equiv k_2 \equiv \frac{1}{2}(\Sigma_{i} K_{i})^{-1.2}
\end{equation}
Equation \eqref{E:18} holds valid for two nearly degenerate heavy Majorana masses and in most general cases $5 \leq K_{i} \leq 100$. To calculate the baryon asymmetry parameter, we have used formula \eqref{E:16}. The results are shown as a variation of the effective neutrino masses calculated for momentum dependent contributions and heavy right handed neutrino contribution as described below. The variations are shown for different values of $M_{W_R}$ and also variations are shown for different strengths of the seesaw mass terms.  \\
Before delving into the results showing the connection between lepton asymmetry and $0\nu\beta\beta$, we have first determined the value of $\eta_{B}$ corresponding to a range of values of $M_{W_R}$ for both normal and inverted hierarchy and also for $X=0.3,0.5,0.7$ as shown in figures \ref{f:1a} and \ref{f:1b}.
\begin{figure}[H]
	\centering
	\includegraphics[scale=0.4]{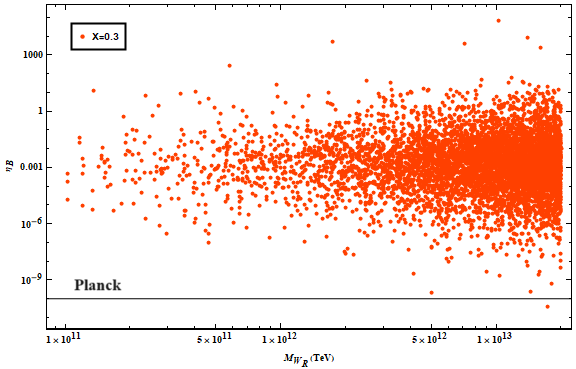}
	\includegraphics[scale=0.4]{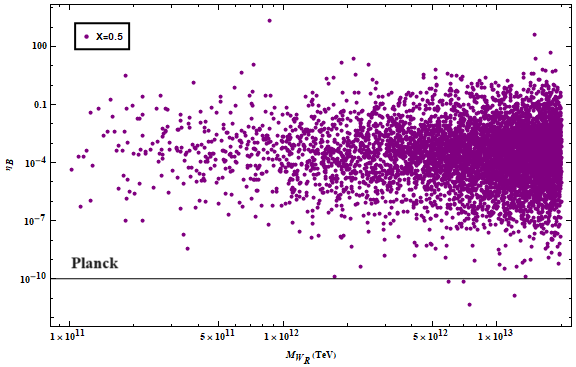}
\end{figure}
Talking about neutrinoless double beta decay $(0\nu\beta\beta)$, in this work, the momentum dependent mechanisms namely, $\lambda$ and $\eta$ contributions have been taken into consideration, which are given by the following formulae,
\begin{itemize}
	\item For $\lambda$ contribution, the dimensionless parameter $\eta_\lambda$ is given by,
	\begin{equation}
		\label{E:19}
		|{\eta_\lambda}| = \Biggl(\frac{M_{W_L}}{M_{W_R}}\Biggl)^{2}|\Sigma_{i}U_{ei}T_{ei}^*|
	\end{equation}
\item For $\eta$ contribution, the dimensionless parameter describing $0\nu\beta\beta$ is given by,
\begin{equation}
	\label{E:20}
	|\eta_{\eta}| = \tan \xi |\Sigma_{i}U_{ei}T_{ei}^*|
\end{equation}
\end{itemize}
In the above equations, $U_{ei}$ represents the first row of the neutrino mixing matrix. $|\Sigma_{i}U_{ei}T_{ei}^*|$ can be simplified to the form $-[M_{D}M_{RR}^{-1}]_{ee}$ as described in \cite{Barry:2013xxa}. T is represented by the equation \eqref{E:28} and,
\begin{equation}
	\label{E:21}
	tan 2\xi = -\frac{2k_1k_2}{v_{R}^2 - v_{L}^2}
\end{equation}
The heavy right-handed (RH) neutrino contribution for $0\nu\beta\beta$ is given by,
\begin{equation}
	\label{E:22}
	M_{eff}^{N} = p^2\Biggl(\frac{M_{W_{L}}^4}{M_{W_{R}}^4}\Biggl)\frac{U_{Rei}^2}{M_i}
\end{equation}
where, $<p^2> = m_e m_p \frac{M_N}{M_\nu}$ is the typical momentum exchange of the process. $m_e$ and $m_p$ are the masses of the electron and proton repsectively and $M_N$ is the nuclear matrix element (NME) for the right-handed neutrino exchange. The allowed value of $p$ is in the range $(100-200)$ MeV. But for our analysis we have used $p = 180 MeV$ \cite{Chakrabortty:2012mh}. Figure \ref{f:1},\ref{f:2} and \ref{f:3} below shows the variation of baryon asymmetry parameter with the effective mass for heavy right-handed neutrino contribution for right-handed gauge boson masses 3,5,10 and 18TeV for type-II seesaw mass term contributing $30\%,50\%$ and $70\%$ to light neutrino mass respectively.

\begin{figure}[H]
	\centering
	\includegraphics[scale=0.4]{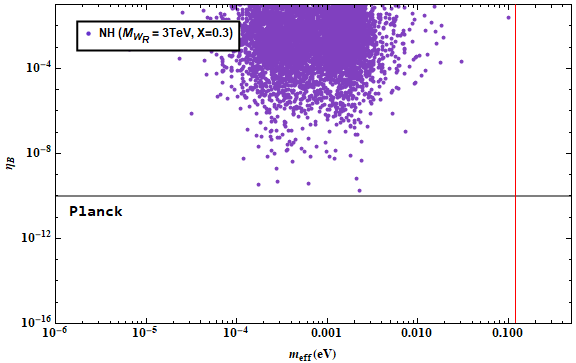}
	\includegraphics[scale=0.4]{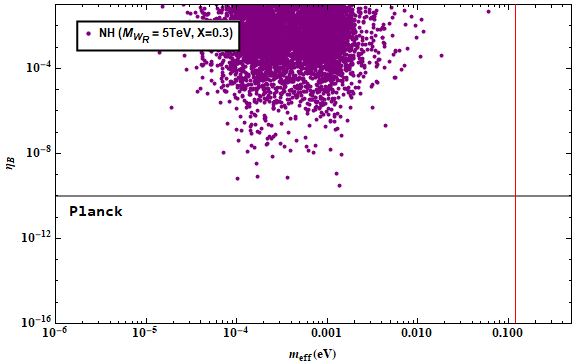}
	\includegraphics[scale=0.4]{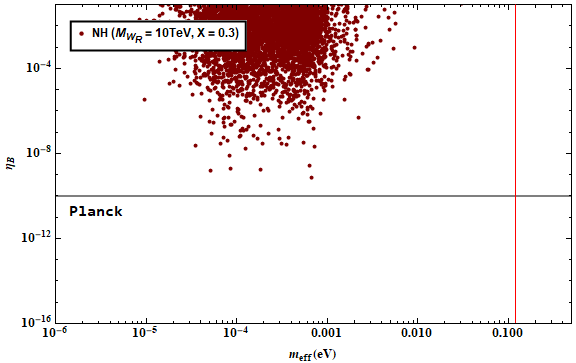}
	\includegraphics[scale=0.4]{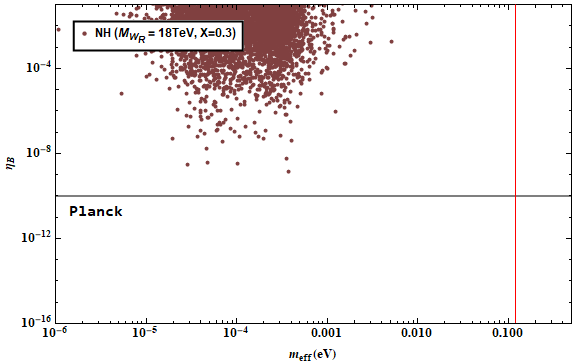}
	\includegraphics[scale=0.4]{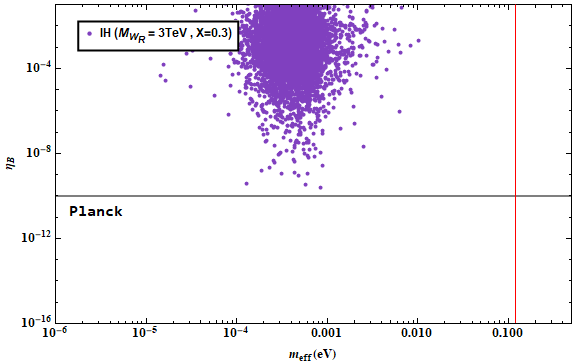}
	\includegraphics[scale=0.4]{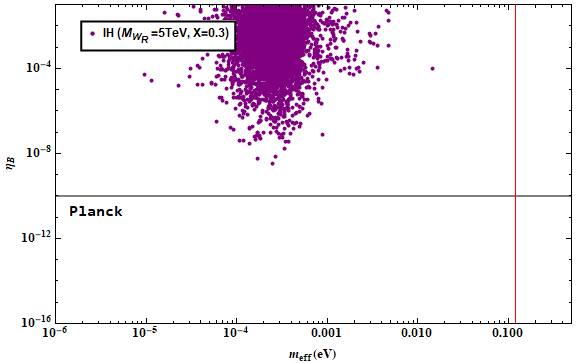}
\end{figure}
\begin{figure}[H]
	\includegraphics[scale=0.4]{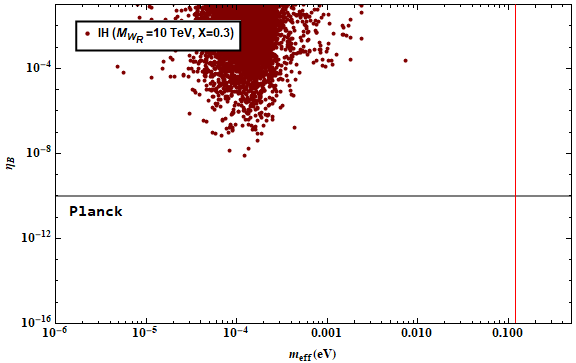}
	\includegraphics[scale=0.4]{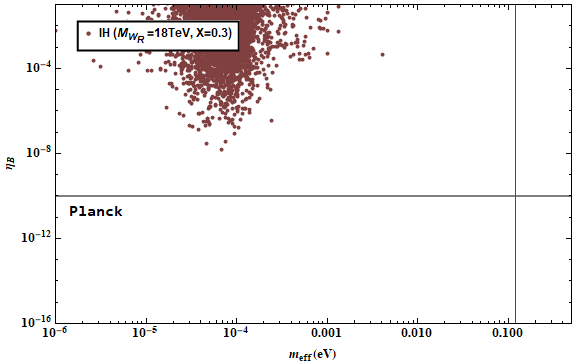}
	\caption{\label{f:1}Variation of $\eta_{B}$ with effective masses corresponding to heavy right-handed neutrino contribution for $M_{W_R}$ = 3,5,10 and 18TeV respectively for normal hierarchy(NH) when $X=0.3$.}
\end{figure}
\begin{figure}[H]
	\centering
	\includegraphics[scale=0.4]{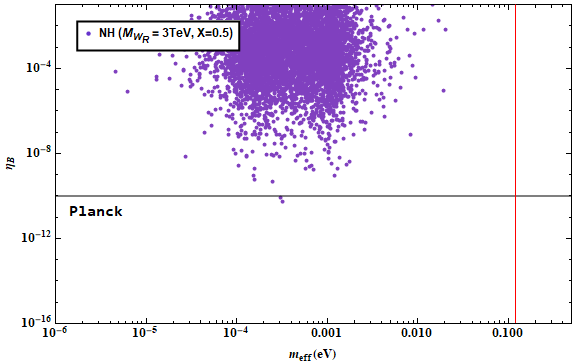}
	\includegraphics[scale=0.4]{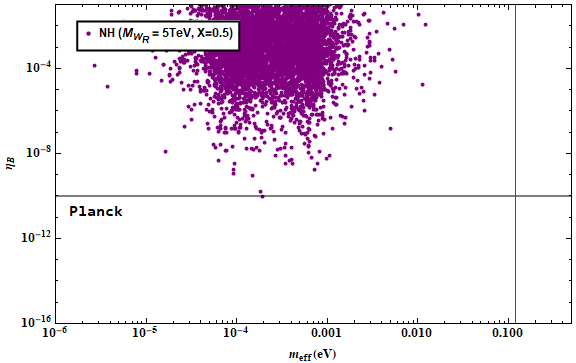}
	\includegraphics[scale=0.4]{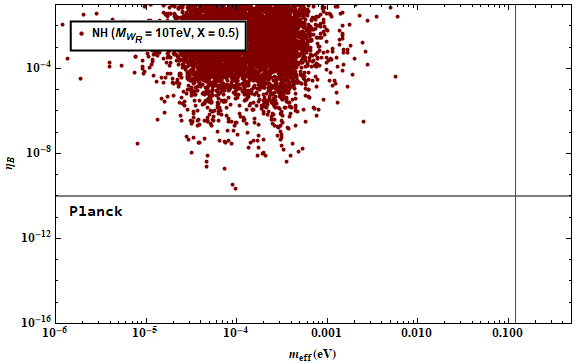}
	\includegraphics[scale=0.4]{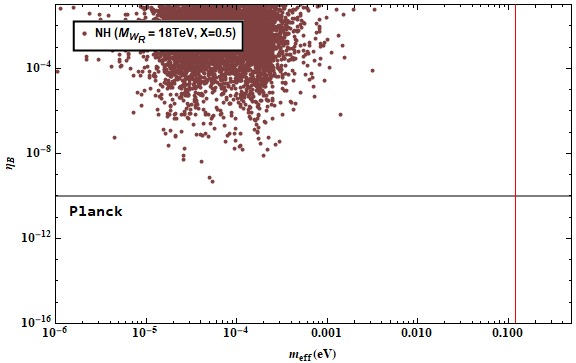}
\end{figure}
\begin{figure}
	\includegraphics[scale=0.4]{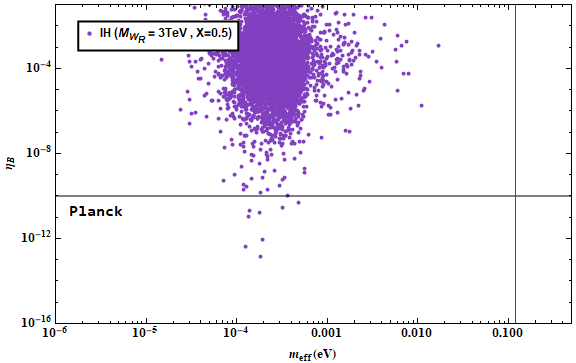}
	\includegraphics[scale=0.4]{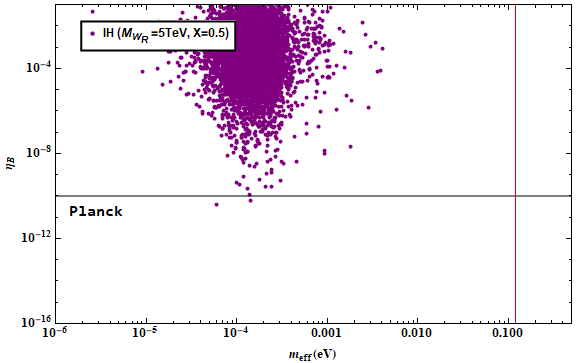}
	\includegraphics[scale=0.4]{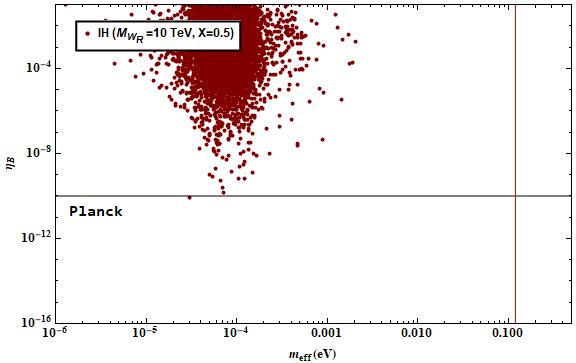}
	\includegraphics[scale=0.4]{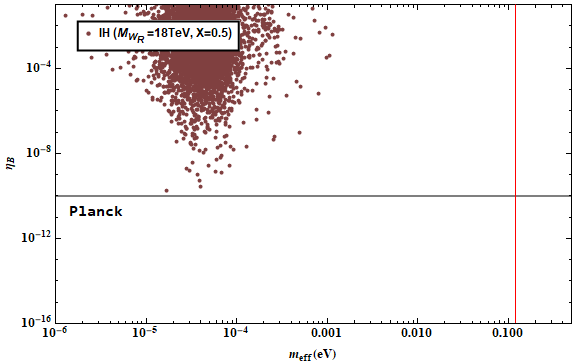}
	\caption{\label{f:2}Variation of $\eta_{B}$ with effective masses corresponding to heavy right-handed neutrino contribution for $M_{W_R}$ = 3,5,10 and 18TeV respectively for NH and IH when $X=0.5$.}
\end{figure}
\begin{figure}[H]
	\centering
	\includegraphics[scale=0.4]{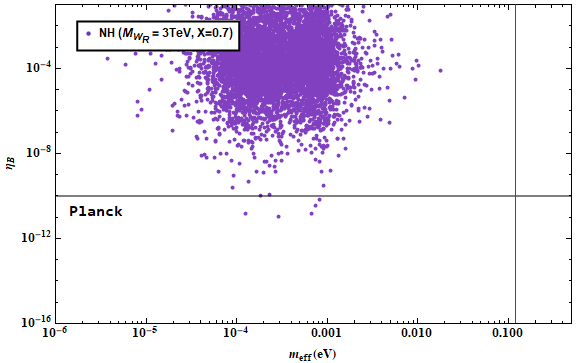}
	\includegraphics[scale=0.4]{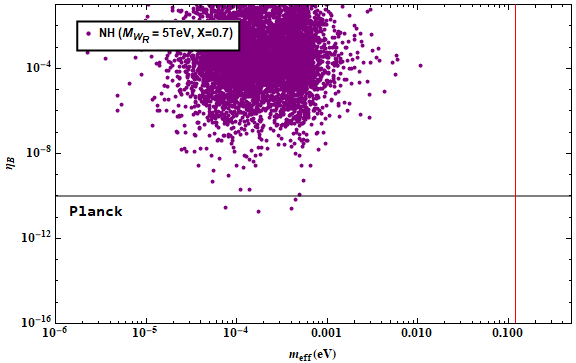}
	\end{figure}
\begin{figure}[H]
	\includegraphics[scale=0.4]{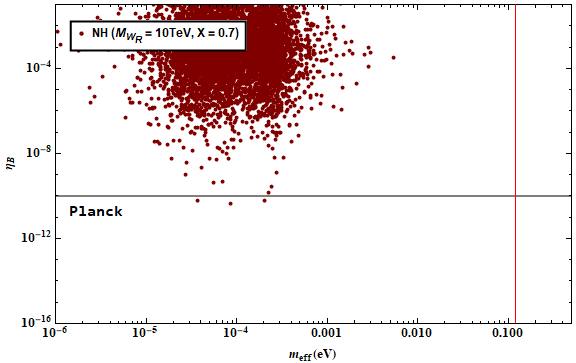}
	\includegraphics[scale=0.4]{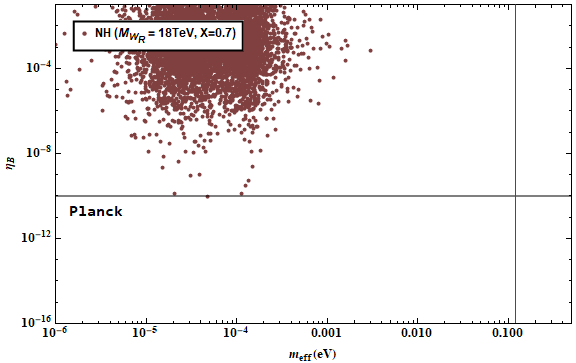}
	\includegraphics[scale=0.4]{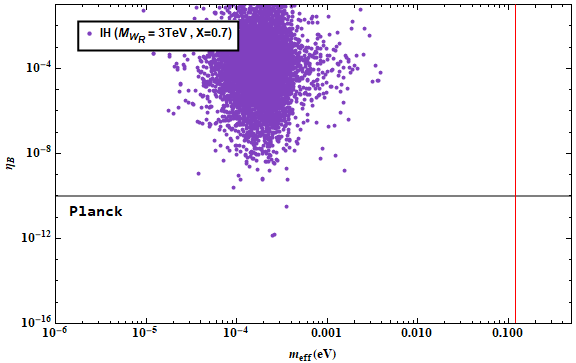}
	\includegraphics[scale=0.4]{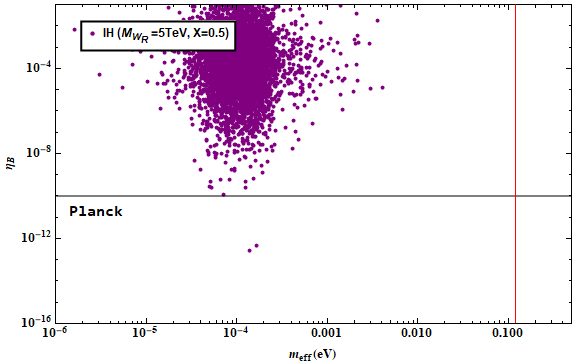}
	\includegraphics[scale=0.4]{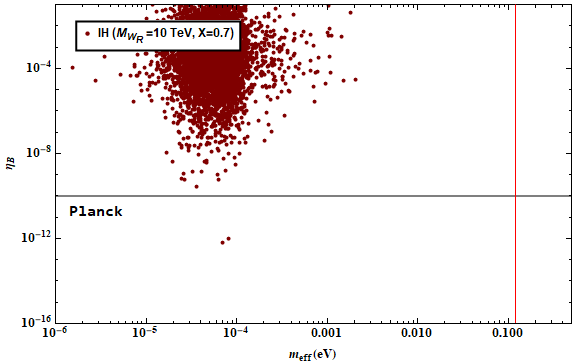}
	\includegraphics[scale=0.4]{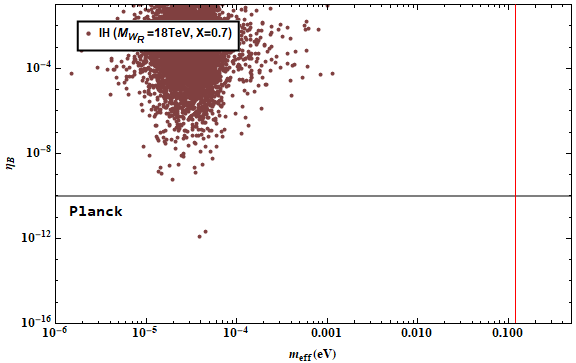}
	\caption{\label{f:3}Variation of $\eta_{B}$ with effective masses corresponding to heavy right-handed neutrino contribution for $M_{W_R}$ = 3,5,10 and 18TeV respectively for NH and IH when $X=0.7$.}
\end{figure}
\begin{figure}[H]
	\centering
	\includegraphics[scale=0.35]{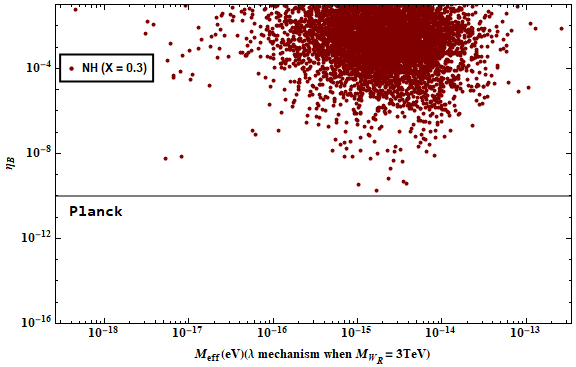}
	\includegraphics[scale=0.35]{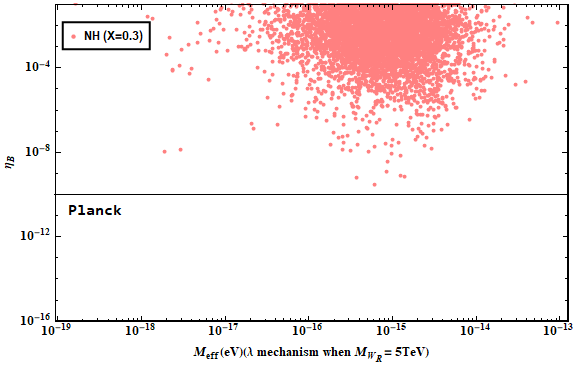}
	\includegraphics[scale=0.35]{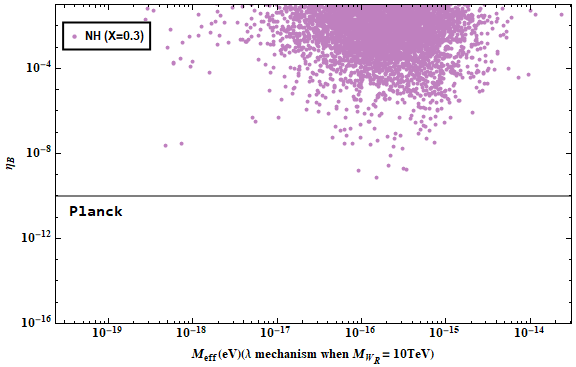}
	\includegraphics[scale=0.35]{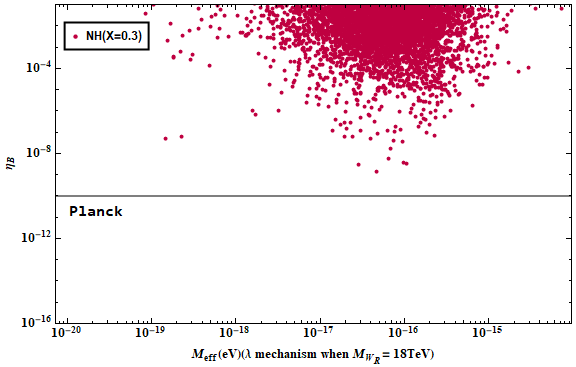}
	\includegraphics[scale=0.35]{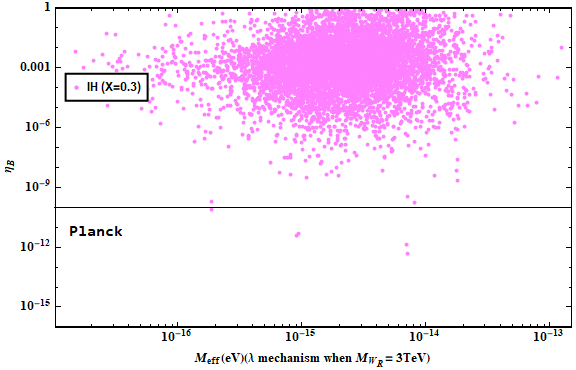}
	\includegraphics[scale=0.35]{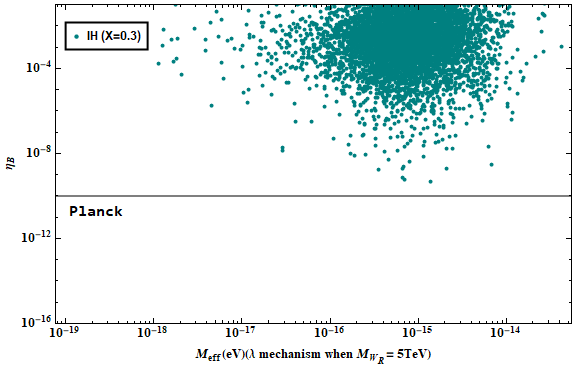}
	\includegraphics[scale=0.35]{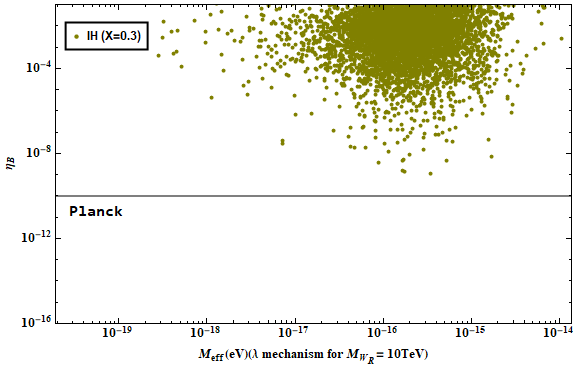}
	\includegraphics[scale=0.35]{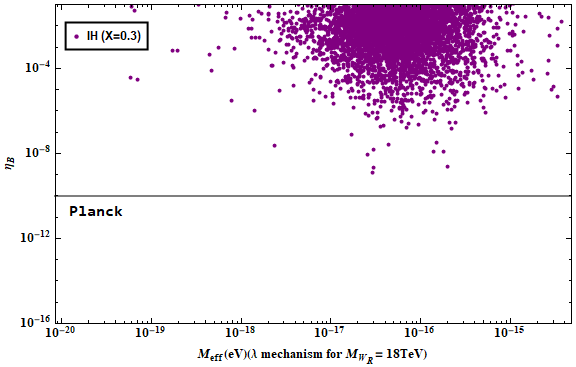}
	\caption{\label{f:4} Variation of effective mass with baryon asymmetry parameter for NH and IH and X=0.3.}
\end{figure}
\begin{figure}[H]
	\centering
	\includegraphics[scale=0.35]{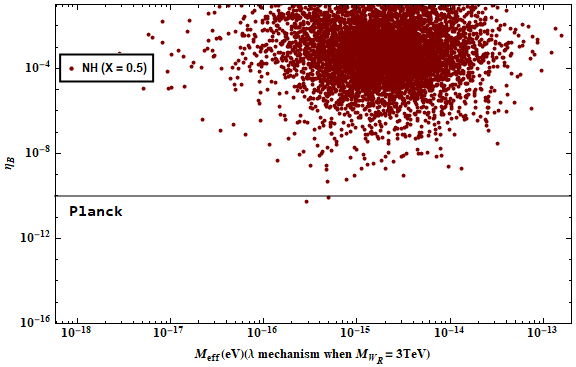}
	\includegraphics[scale=0.35]{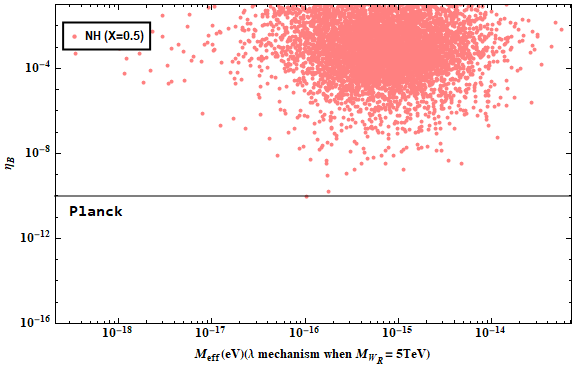}
	\includegraphics[scale=0.35]{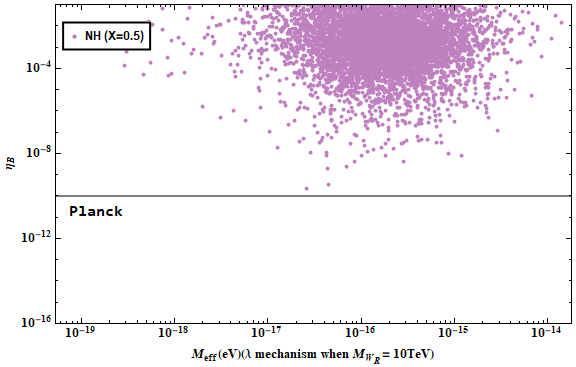}
	\includegraphics[scale=0.35]{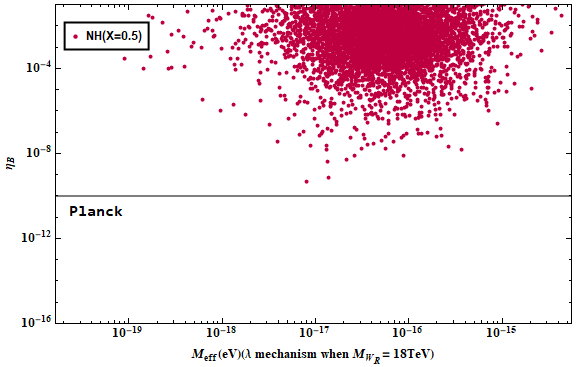}
	\includegraphics[scale=0.35]{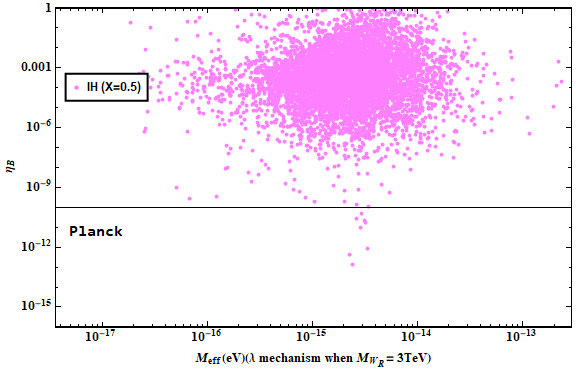}
	\includegraphics[scale=0.35]{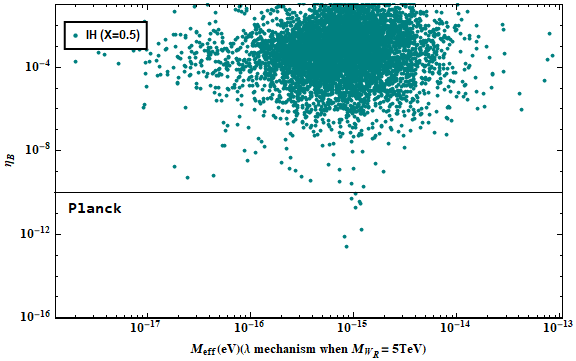}
	\includegraphics[scale=0.35]{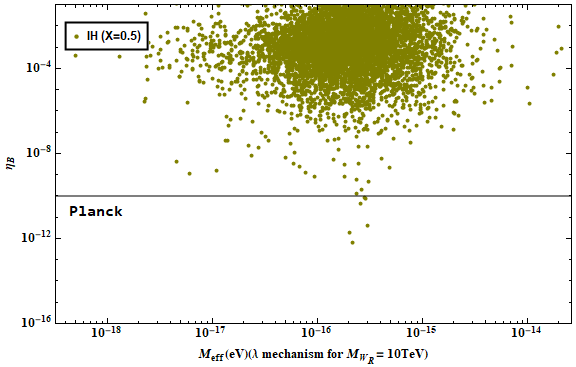}
	\includegraphics[scale=0.35]{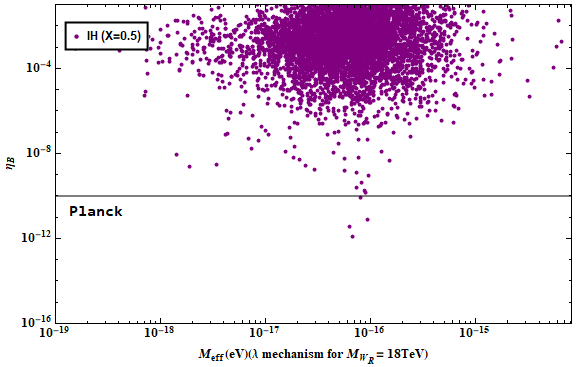}
	\caption{\label{f:5} Variation of effective mass with baryon asymmetry parameter for NH and IH and X=0.5.}
\end{figure}
\begin{figure}[H]
	\centering
	\includegraphics[scale=0.35]{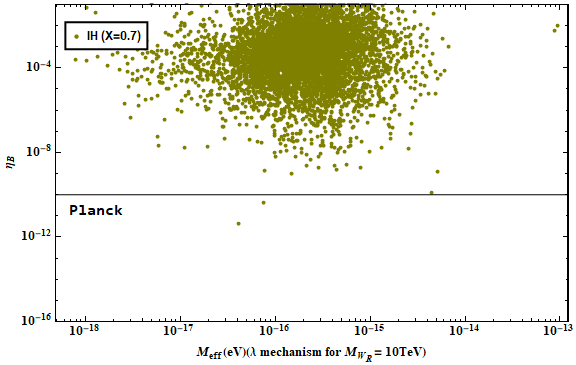}
	\includegraphics[scale=0.35]{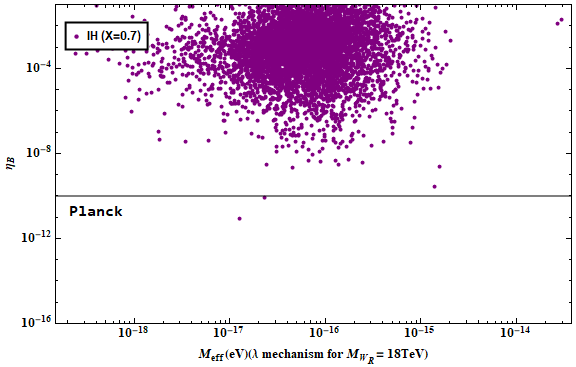}
	\caption{\label{f:6} Variation of effective mass with baryon asymmetry parameter for NH and IH and X=0.7.}
\end{figure}

The effective mass for $\lambda$ contribution, that is, $|m_{eff}^{\lambda}|$  is expressed in terms of the dimensionless parameter $\eta_{\lambda}$ as,
\begin{equation}
	\label{E:23}
	|\eta_{\lambda}| = \frac{|m_{eff}^{\lambda}|}{m_e}
\end{equation}
The variations of baryon asymmetry parameter with the effective masses corresponding to the aforementioned contribution are shown in figures \ref{f:4} to \ref{f:6}.
\begin{figure}[H]
	\centering
	\includegraphics[scale=0.35]{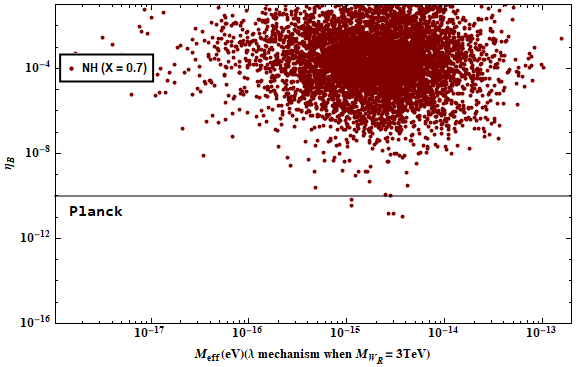}
	\includegraphics[scale=0.35]{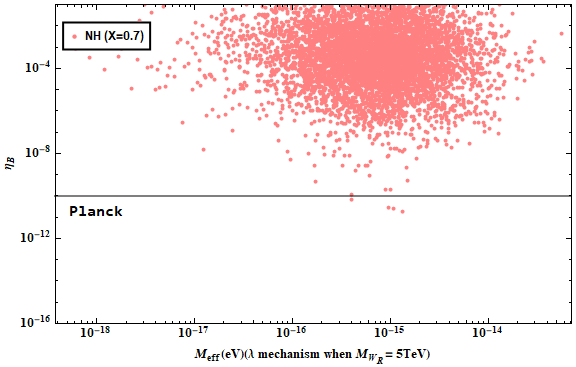}
	\includegraphics[scale=0.35]{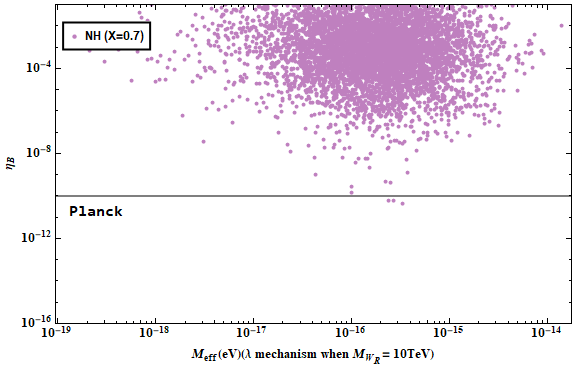}
	\includegraphics[scale=0.35]{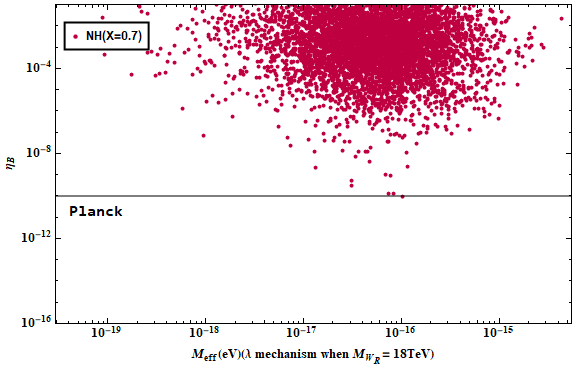}
	\includegraphics[scale=0.35]{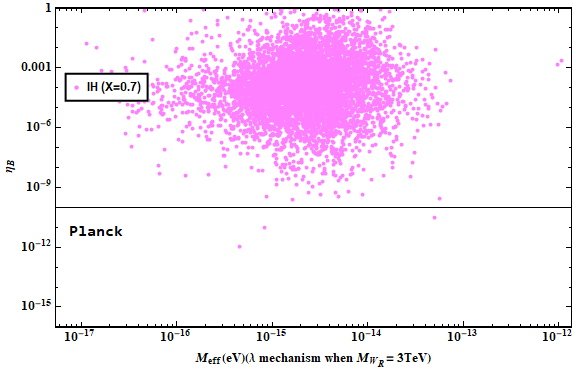}
	\includegraphics[scale=0.35]{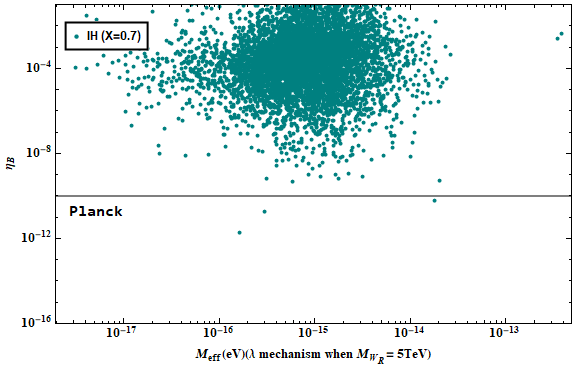}
\end{figure}
The effective mass for $\eta$ contribution, that is, $|m_{eff}^{\eta}|$  is expressed in terms of the dimensionless parameter $\eta_{\eta}$ as,
\begin{equation}
	\label{E:24}
	|\eta_{\eta}| = \frac{|m_{eff}^{\eta}|}{m_e}
\end{equation}
\begin{figure}[H]
	\centering
	\includegraphics[scale=0.35]{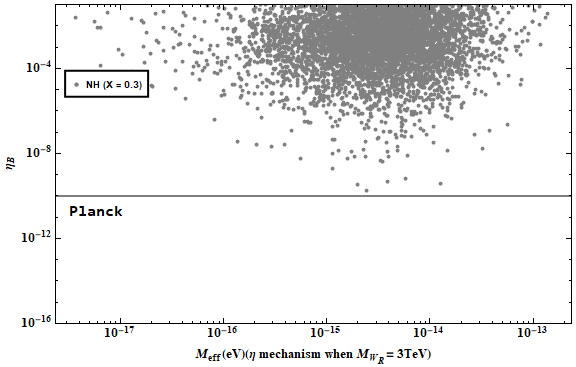}
	\includegraphics[scale=0.35]{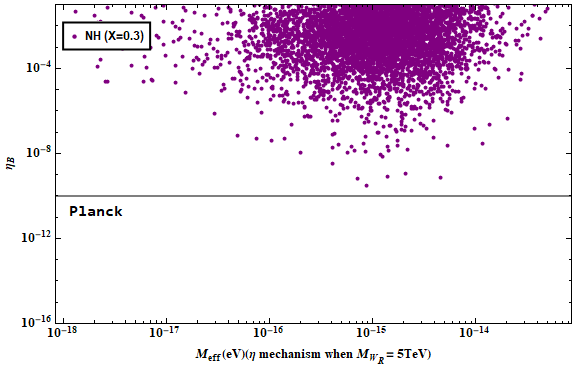}
	\includegraphics[scale=0.35]{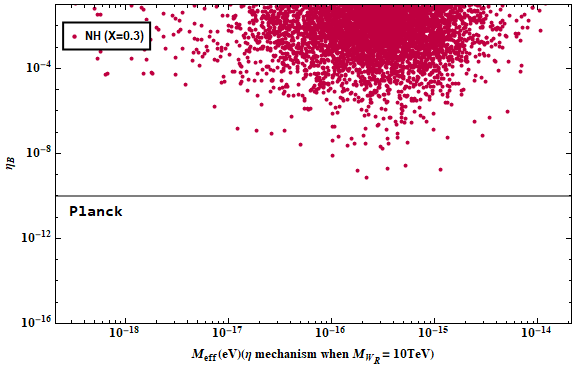}
	\includegraphics[scale=0.35]{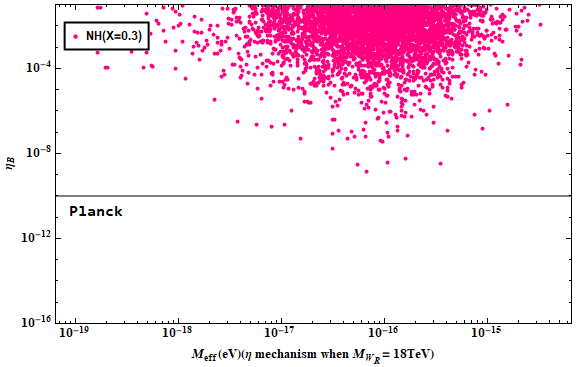}
	\includegraphics[scale=0.35]{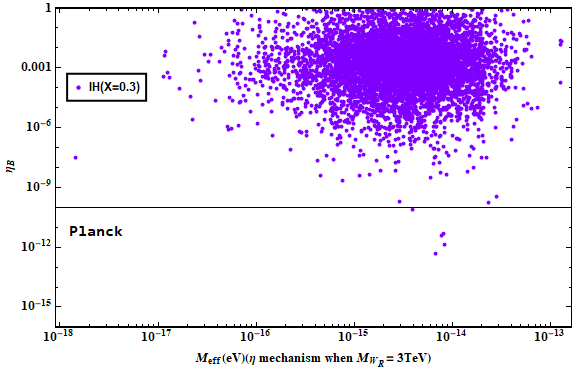}
	\includegraphics[scale=0.35]{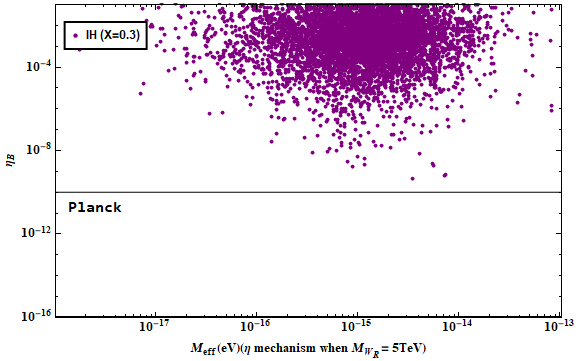}
	\includegraphics[scale=0.35]{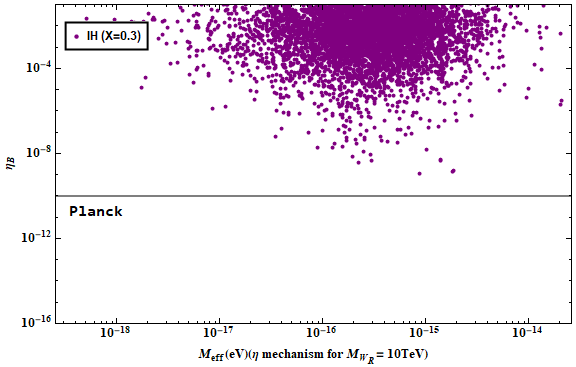}
	\includegraphics[scale=0.35]{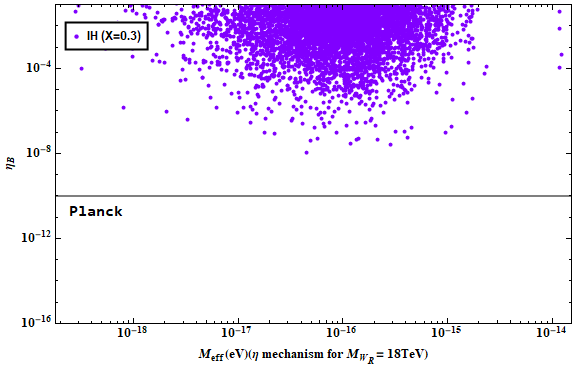}
	\caption{\label{f:7} Variation of effective mass with baryon asymmetry parameter for NH and IH and X=0.3.}
\end{figure}
\begin{figure}[H]
	\centering
	\includegraphics[scale=0.35]{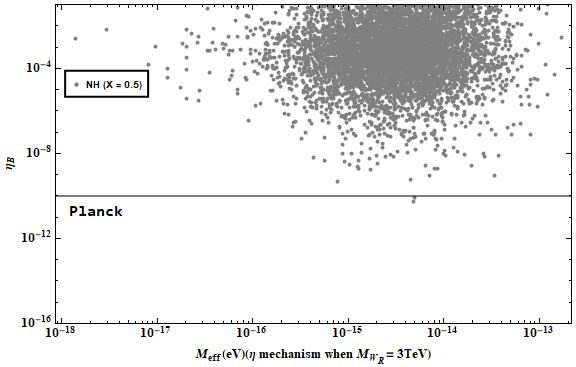}
	\includegraphics[scale=0.35]{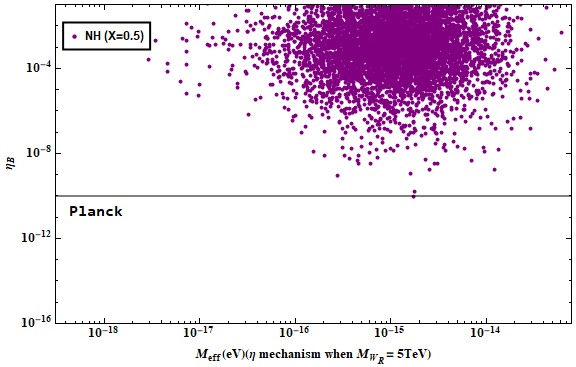}
	\includegraphics[scale=0.35]{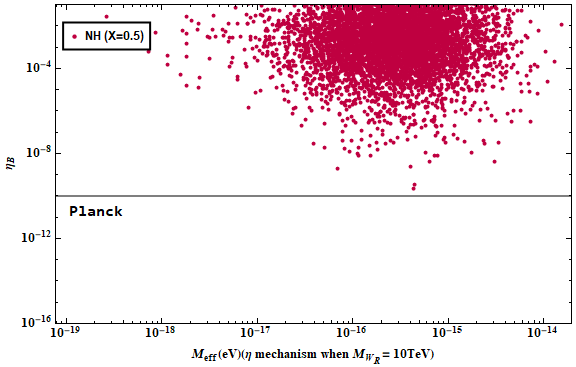}
	\includegraphics[scale=0.35]{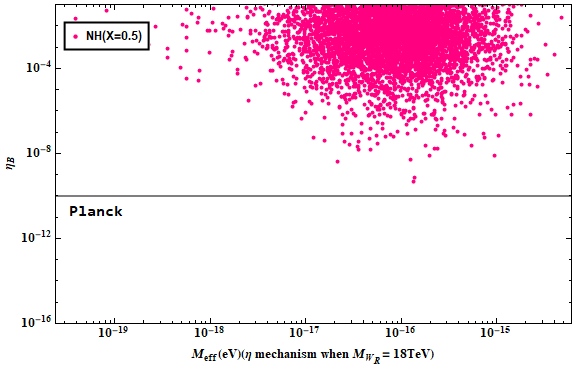}
	\includegraphics[scale=0.35]{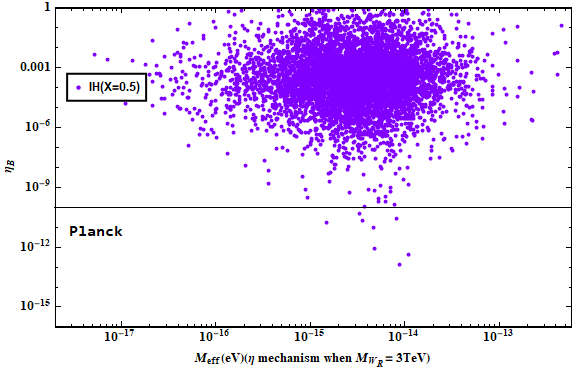}
	\includegraphics[scale=0.35]{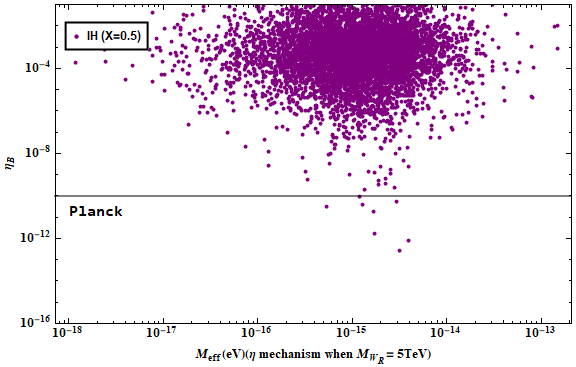}
	\includegraphics[scale=0.35]{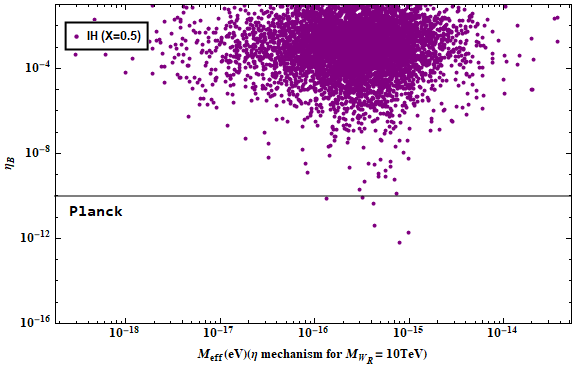}
	\includegraphics[scale=0.35]{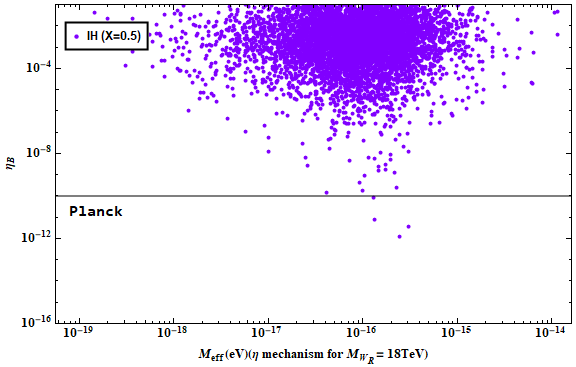}
	\caption{\label{f:8} Variation of effective mass with baryon asymmetry parameter for NH and IH and X=0.5.}
\end{figure}
\begin{figure}[H]
	\centering
	\includegraphics[scale=0.32]{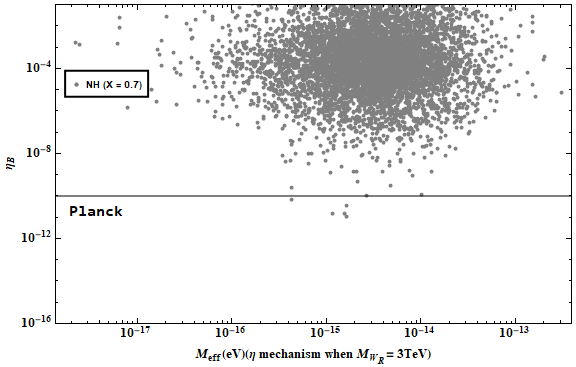}
	\includegraphics[scale=0.32]{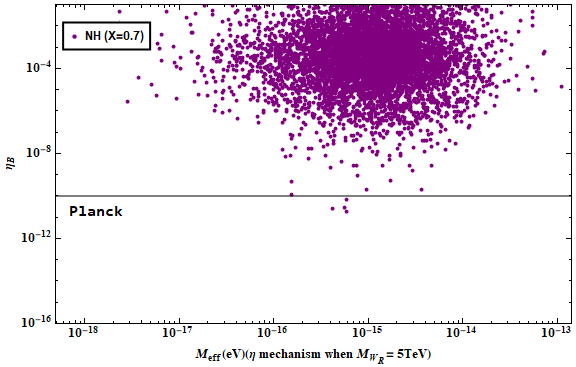}
	\includegraphics[scale=0.32]{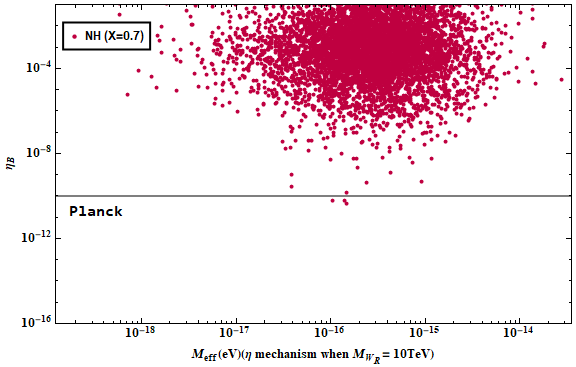}
	\includegraphics[scale=0.32]{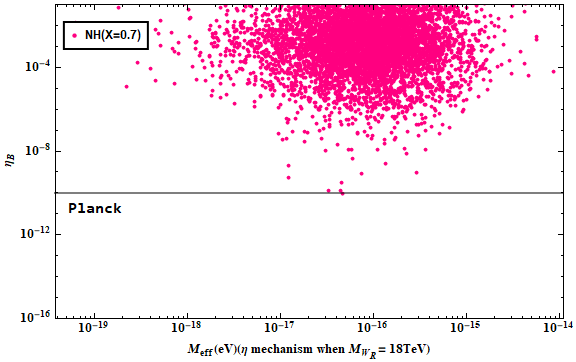}
	\includegraphics[scale=0.32]{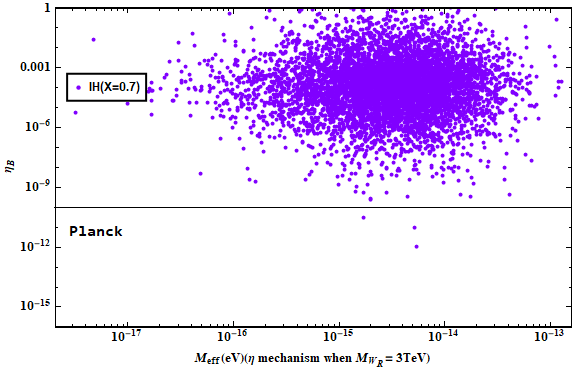}
	\includegraphics[scale=0.32]{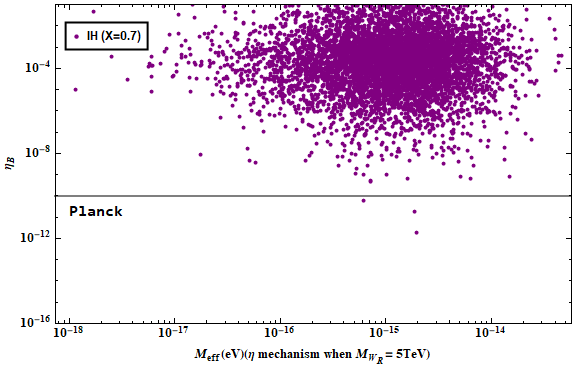}
	\includegraphics[scale=0.32]{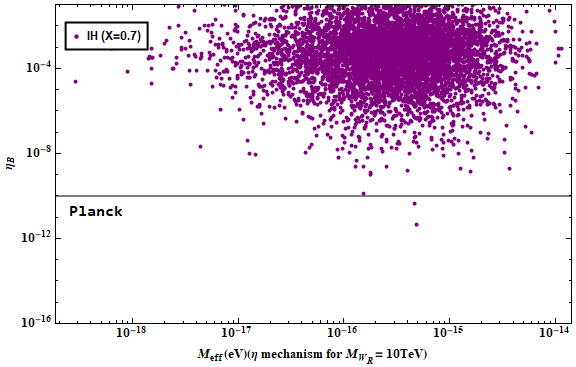}
	\includegraphics[scale=0.32]{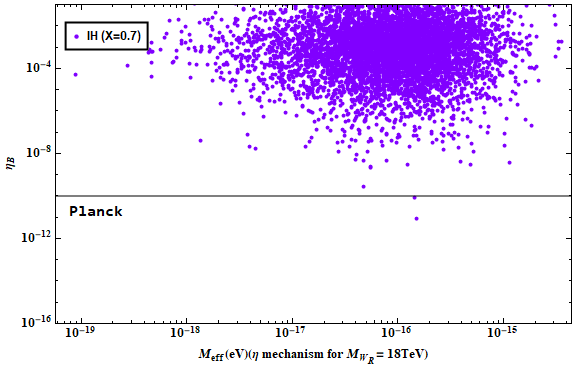}
	\caption{\label{f:9} Variation of effective mass with baryon asymmetry parameter for NH and IH and X=0.7.}
\end{figure}
Figures \ref{f:7} to \ref{f:9} shows the variation of baryon asymmetry parameter with the effective masses for $\eta$ contribution of $0\nu\beta\beta$.
\begin{figure}[H]
	\centering
	\includegraphics[scale=0.3]{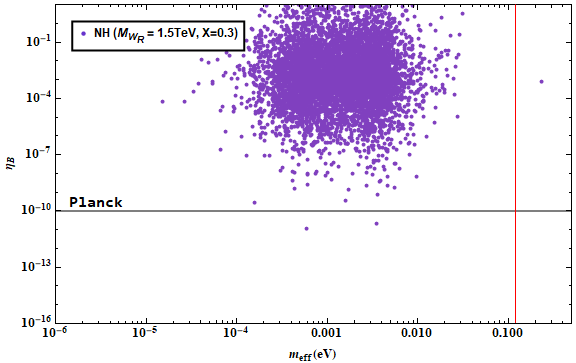}
	\includegraphics[scale=0.3]{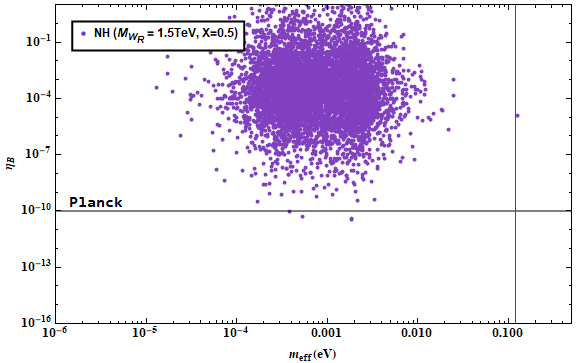}
	\includegraphics[scale=0.3]{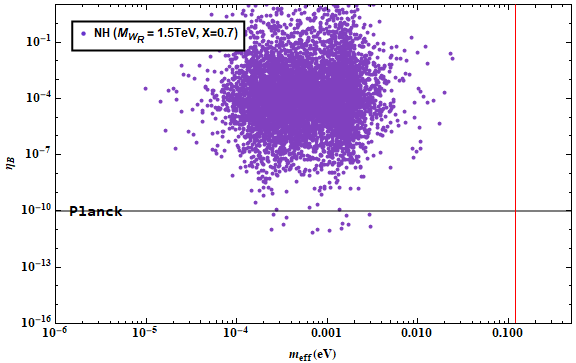}
	\includegraphics[scale=0.3]{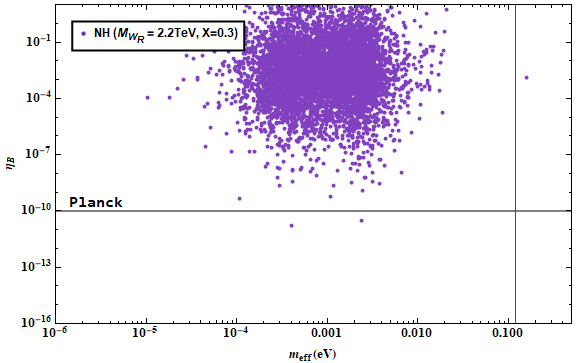}
	\includegraphics[scale=0.3]{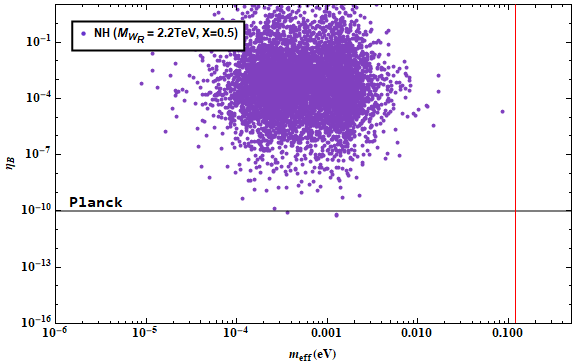}
	\includegraphics[scale=0.3]{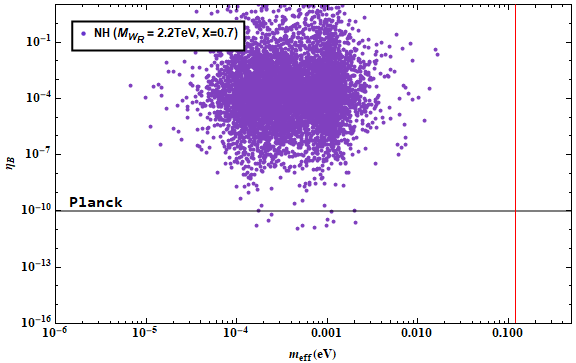}
	\includegraphics[scale=0.3]{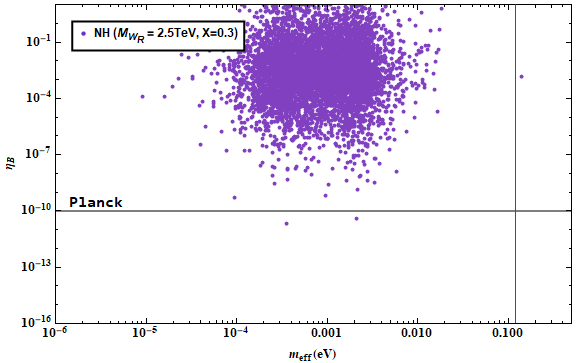}
	\includegraphics[scale=0.3]{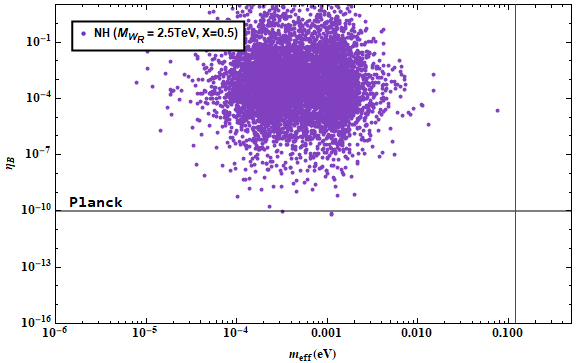}
	\includegraphics[scale=0.3]{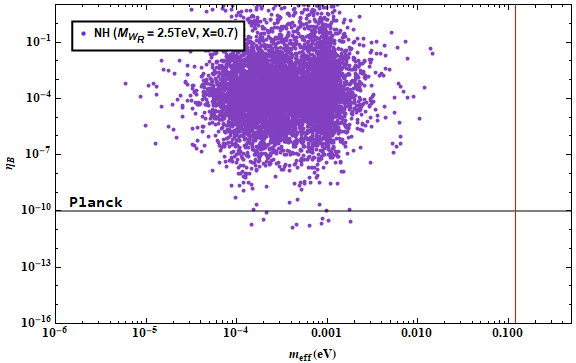}
	\caption{\label{f:10} Variation of effective mass with baryon asymmetry parameter for NH and X=0.3,0.5 and 0.7.}
\end{figure}

Figure \ref{f:10} shows the variation of baryon aymmetry parameter with the effective mass corresponding to right-handed neutrino contribution for $1.5,2.2$ and $2.5 TeV$ for normal hierarchy.
\begin{figure}[H]
	\centering
	\includegraphics[scale=0.35]{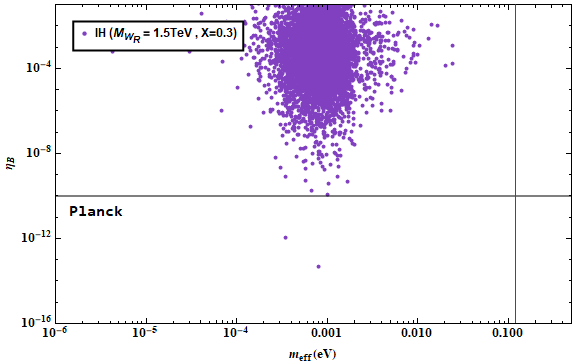}
	\includegraphics[scale=0.35]{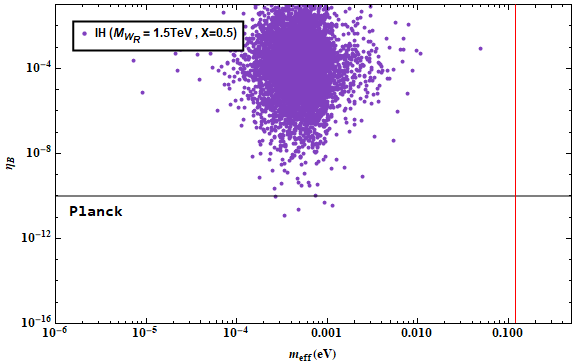}
	\includegraphics[scale=0.35]{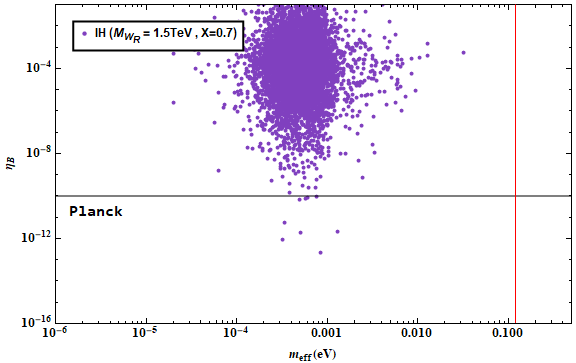}
	\includegraphics[scale=0.35]{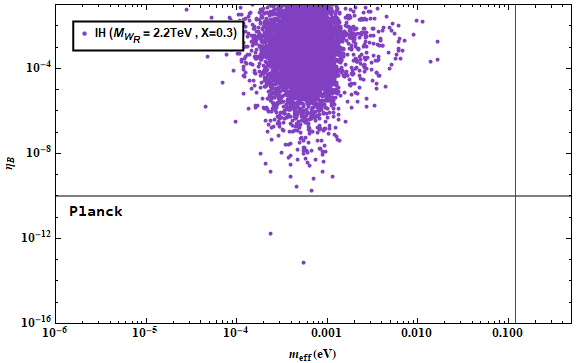}
	\includegraphics[scale=0.35]{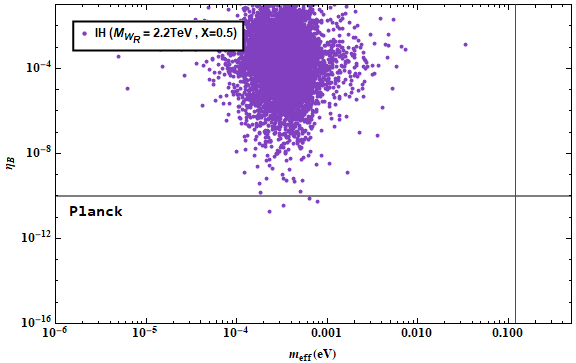}
	\includegraphics[scale=0.35]{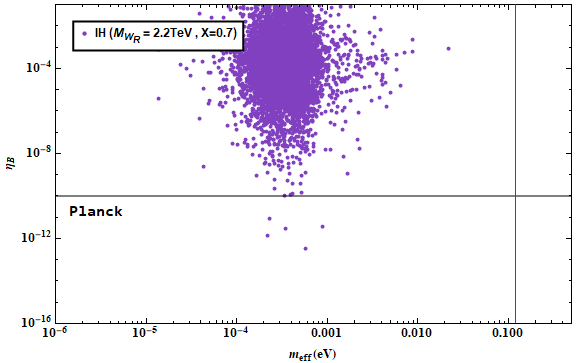}
	\includegraphics[scale=0.35]{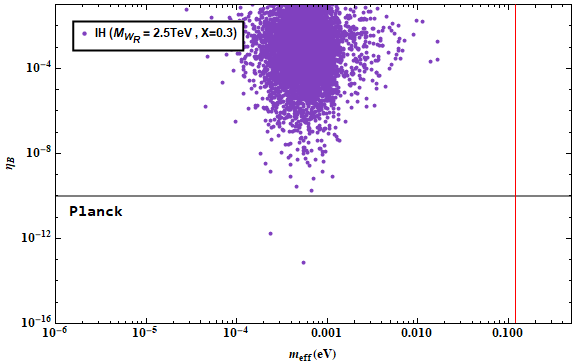}
	\includegraphics[scale=0.35]{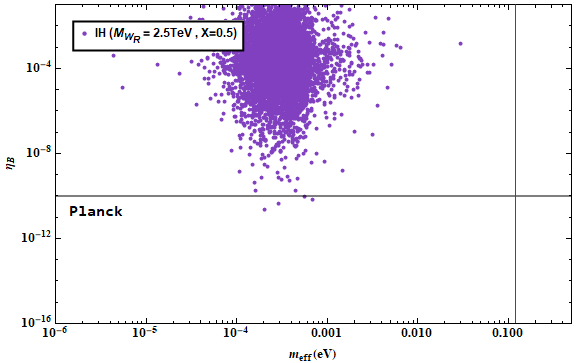}
\end{figure}
\begin{figure}[H]
	\centering
\includegraphics[scale=0.35]{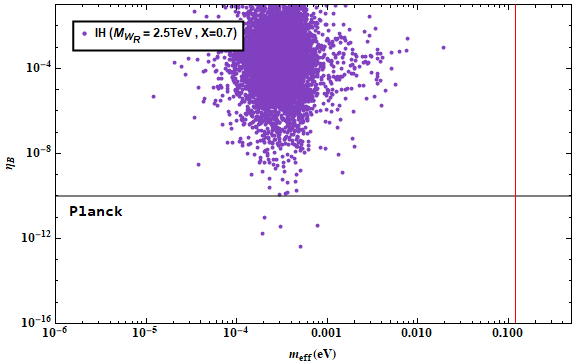}
\caption{\label{f:11} Variation of effective mass for heavy right-handed neutrino contribution with baryon asymmetry parameter for IH and X=0.3,0.5 and 0.7.}
\end{figure}
Figure \ref{f:11} shows the variation of baryon aymmetry parameter with the effective mass corresponding to right-handed neutrino contribution for $1.5,2.2$ and $2.5 TeV$ for inverted hierarchy. It has been observed that the heavy right-handed neutrino corresponding to $X=0.3$ and $X=0.5$ is very less in case of $1.5 , 2.2$ and $2.5$ TeV. But when $X=0.7$ that is, when the type-II seesaw mass term is dominant, there are some amount of data obtained within the experimental limits. Same has been observed in case of lambda and eta contributions of $0\nu\beta\beta$ as well and as such, we are showing only the plots corresponding to $X=0.7$.
\begin{figure}[H]
	\centering
	\includegraphics[scale=0.35]{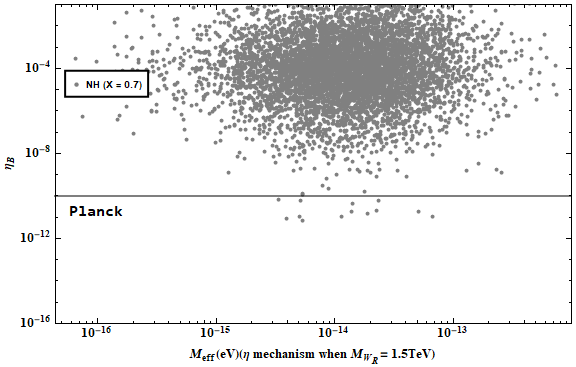}
	\includegraphics[scale=0.35]{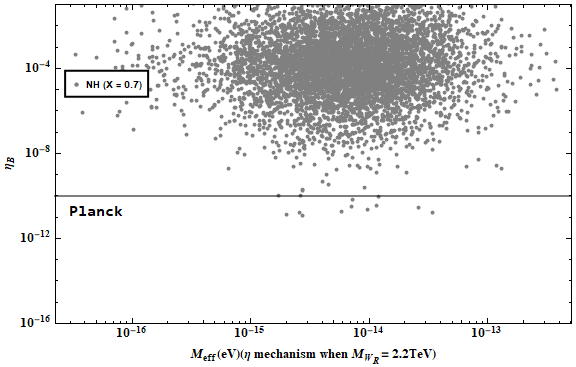}
\end{figure}
\begin{figure}[H]
	\centering
\includegraphics[scale=0.35]{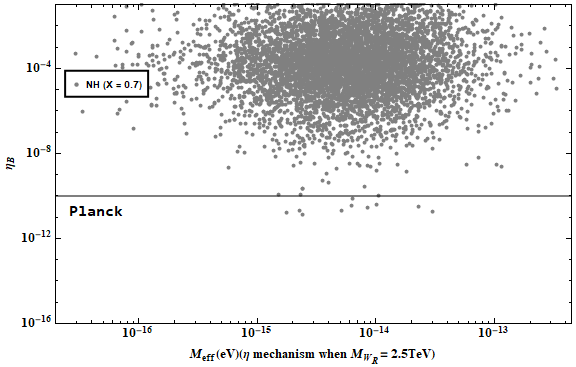}
\caption{\label{f:12} Variation of effective mass for $\eta$ contribution with baryon asymmetry parameter for NH and X=0.7.}
\end{figure}
	
\begin{figure}[H]
	\centering
	\includegraphics[scale=0.35]{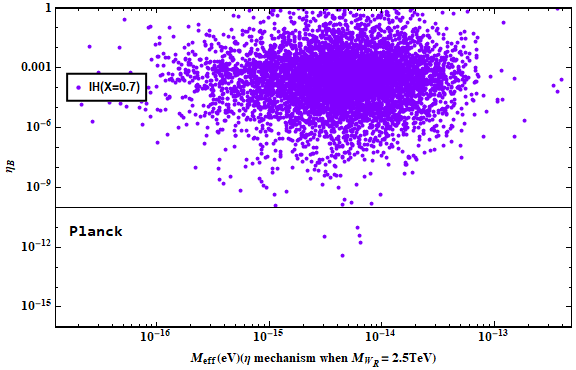}
	\caption{\label{f:13} Variation of effective mass for $\eta$ contribution with baryon asymmetry parameter for IH and X=0.7.}
\end{figure}

\begin{figure}[H]
	\centering
	\includegraphics[scale=0.35]{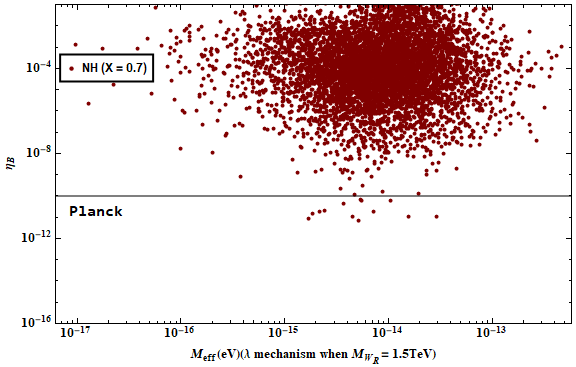}
	\includegraphics[scale=0.35]{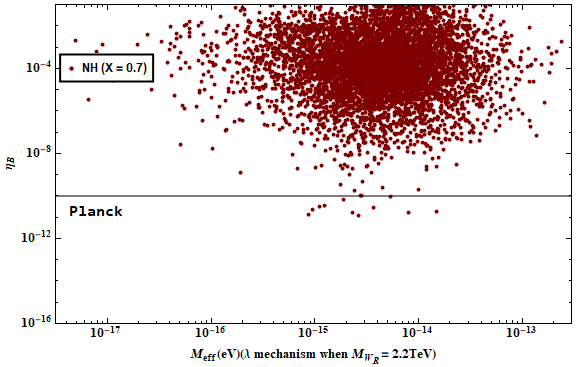}
	\includegraphics[scale=0.35]{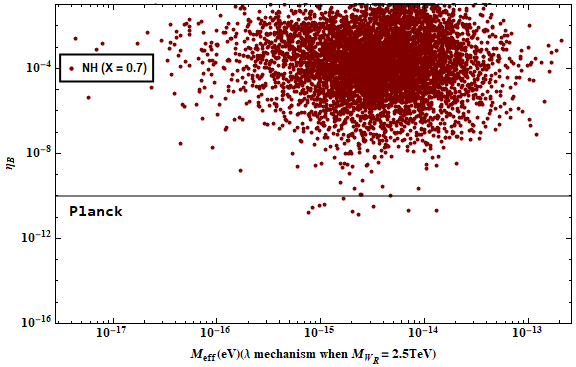}
	\caption{\label{f:14} Variation of effective mass for $\lambda$ contribution with baryon asymmetry parameter for NH and X=0.7.}
\end{figure}

\begin{figure}[H]
	\centering
	\includegraphics[scale=0.35]{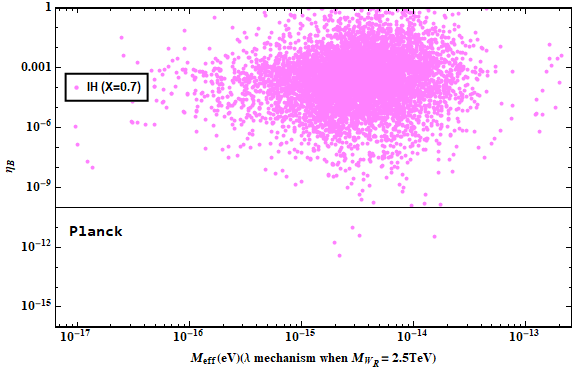}
	\caption{\label{f:15} Variation of effective mass for $\lambda$ contribution with baryon asymmetry parameter for IH and X=0.7.}
\end{figure}

Figure \ref{f:12} shows the variation of effective mass corresponding to $\eta$ contribution with the baryon asymmetry parameter for the type-II seesaw mass contributing $70\%$ towards the neutrino mass. The plots are for normal hierarchy. The figure \ref{f:13} shows the variations for inverted hierarchy.
\begin{figure}[H]
	\centering
	\includegraphics[scale=0.35]{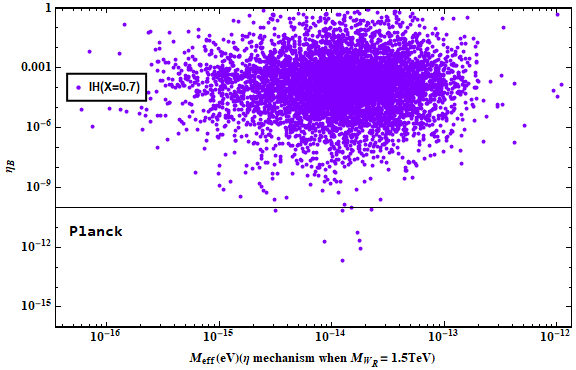}
	\includegraphics[scale=0.35]{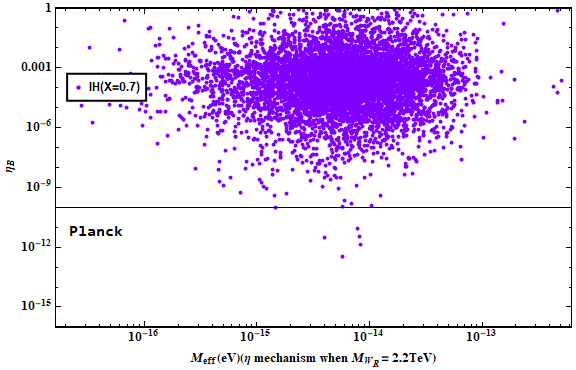}
\end{figure}
Variations of baryon asymmetry parameter with effective mass corresponding to $\lambda$ contribution have also been shown in figures \ref{f:14} and \ref{f:15}.
\begin{figure}[H]
	\centering
	\includegraphics[scale=0.35]{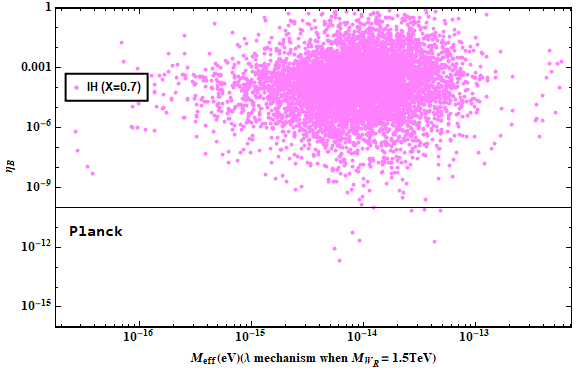}
	\includegraphics[scale=0.35]{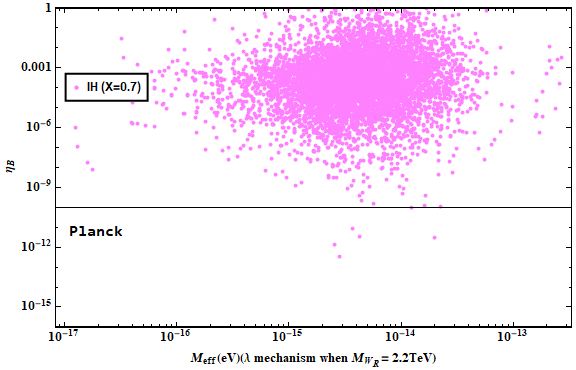}
\end{figure}
\begin{center}
	\section{\label{lrsm6}Discussion}
\end{center}

The following points describe the complete study and analysis of the present work and the results obtained hereafter,
\begin{itemize}
	\item Left-Right Symmetric model has been realized with the help of modular group of level 3 and weight 2 ($\Gamma(3)$), isomorphic to the $A_4$ non-abelian discrete symmetry group, as discussed in the appendix.
	\item  The aim of the study focusses on describing the variation of baryogenesis (via leptogenesis) with variations in the value of the right-handed gauge boson mass and strength of the type-II seesaw mass.
	\item In several previous literatures, it was stated that while taking into consideration resonant leptogenesis, the dominant contribution was given by type-I seesaw mass, however, we have also taken into consideration the strength of the type-II seesaw mass term and it has been found that the analysis gives considerable results for the same.
	\item We have considered strength of the type-II seesaw mass to be 30\% , 50\% and 70\%, and masses of the right-handed gauge bosons are considered to be $M_{W_R} = 1.5,2.2,2.5,3,5,10,18$ TeV. Here, the condition $M_N \leq M_{W_R}$ has been used.
	\item From the present analysis, it can be stated that a lower bound of $M_{W_R}$ $\geq$ 1.5 TeV has been obatined for the case $g_{L} = g_{R}$, that is, when the gauge couplings corresponding to left-handed and right-handed components are equal. In the present work, the Yukawa couplings have been expressed in terms of modular forms $(Y_1,Y_2,Y_3)$ and the values determined for the same have been found to be much larger than the canonical value of $O(10^{-5.5}$) \cite{Dev:2015cxa,BhupalDev:2014hro}. Due to the enhanced Yukawa couplings, the reaction rates of the two body decay becomes more dominant over the $W_R$ mediated scatterings as well as the three body decays \cite{BhupalDev:2014hro}. In the present model, the CP asymmetry parameter has also been numerically investigated rather than considering it as one.\\
		As stated in \cite{BhupalDev:2015khe}, we can take into consideration two different ways of deriving a lower bound on the value of $M_{W_R}$. The first on being the careful scanning of the parameter space of the model and the second one being the use of the approximate expression \ref{e:1} \cite{BhupalDev:2014hro,BhupalDev:2015khe},
		\begin{equation}
			\label{e:1}
			|\eta^{\Delta L}(z_c)|=\frac{9}{4\zeta(3)}\frac{z_c^{2}K_{1}(z_c)}{S_R(z_c)}\frac{r_{s}r_{d}^2}{(3+r_{s}r_{d})(3+r_{s})}\epsilon_{total}^{Y}
		\end{equation}
		where, $r_{s}$ and $r_{d}$ are the relative washout strengths of the two-body decays, $W_R$ mediated decays and scatterings. $\zeta(x)$ is the Riemann function, $K_{1}(z_c)$is the first order Bessel function and $z=\frac{m_{N_1}}{T}$, $T$ being the temperature of the Universe, is a dimensionless variable, and $z_c$ is the critical value beyond which the sphaleron processes become ineffective\cite{BhupalDev:2014hro,BhupalDev:2015khe}. The value of these strengths depend upon the values of $M_{N}$, $M_{W_R}$ and most importantly the Yukawa couplings. we have used the first method to determine the lower bound on $M_{W_R}$, which was found to be $M_{W_R} \geq 1.5 TeV$. As stated in the paper, \cite{BhupalDev:2014hro}, a dependence on the Yukawa couplings of the decay rates can be mathematically expressed as $D \equiv Y^{2}$ which suggests that with larger value of Yukawa couplings, the decay rate will be larger as compared to that of the scatterings induced due to $RH$ currents.
	\item The analysis shows that the weaker bounds on the $M_{W_R}$ mass is applicable only in the case where the type-II term contributes dominantly. This might be the result of the values of the Yukawa couplings and textures of the mass matrices obtained as a result of implementing modular symmetry in LRSM rather than flavor symmetry.
	\item As already stated that, a lower bound has been found through the present study on $M_{W_R}$, but we have also calculated the baryon asymmetry parameter for values less than and also greater than $1.5 TeV$, but the points are more specific only when $M_{W_R} \geq 1.5 TeV$. This lower bound be a result of the larger Yukawa couplings obtained in the present work ranging from $10^{-4}$ to $10^{-1}$ and some maximum values are also greater than $1$.
	\item We have also shown the variations of $\eta_{B}$ with $M_{W_R}$ to emphasize on the fact that even though the range of $M_{W_R}$ is taken to be quite high, but successful explanation of leptogenesis and its connection with the low energy phenomena, that is, Neutrinoless Double Beta Decay can be put forward, only for certain values of X (most preferably $X > 0.5$) that is when the type-II seesaw mass term is dominant and only at specific values of $M_{W_R}$. The results are found to be more reliable for inverted hierarchy as compared to that of normal hierarchy and for the range $1.5 \leq M_{W_R} \leq 2.5$, the values predicted by the model for baryon asymmetry are found to be consistent with the observed value for both normal and inverted hierarchy.
\end{itemize}
Several previous works have been done taking into consideration the flavor symmetric realization of LRSM as studied in \cite{Boruah:2022bvf,Rodejohann:2015hka,Sahu:2020tqe,Boruah:2021ktk} etc. The authors in the mentioned literatures have also carried out phenomenological studies related to the model. However, in the present work we have not considered flavor symmetry but, we have taken into consideration modular symmetry. The use of modular symmetry provides the advatage of maintaining the minimality in construction of the model without any flavons. In the present work, we have considered modular group of level 3, $\Gamma(3)$ and weight 2, which is isomorphic to $A_4$ non-abelian discrete symmetry group. Here, it is to be noted that the present model has been designed in the non-SUSY framework and as such, Left-Right Symmetric Model does not restrict the use of infinite number of modular forms, but as we are using $\Gamma(3)$ modular group having weight 2, as stated in the appendix in table \ref{table:3}, we will have three number of modular forms represented as $(Y_1,Y_2,Y_3)$. As such, the mass matrices corresponding to the present model have been expressed in terms of these three modular forms, discussed in section (III) of the manuscript.\\
Some previous work as dicussed in \cite{Borgohain:2017akh,Borgohain:2017inp} have also taken into account the study of several phenomenology like $0\nu\beta\beta$, Lepton Flavor Violation (LFV) with changing values of $M_{W_R}$. In the present work, we have also taken into account the different values of $M_{W_R}$ to study the relation between $0\nu\beta\beta$ and leptogenesis, however it has been done taking into account different strengths of the type-II seesaw mass term. One of the most significant results of the present work can be stated to be the observation of considerable amount of leptogenesis phenomena taking place when the leading contribution of neutrino mass is given by the type-II seesaw mass term ($70\%$). As stated by the authors in \cite{Borgohain:2017akh} that while taking into consideration the phenomena of resonant leptogenesis, type-I seesaw term plays the dominant role and the contribution of $\Delta_L$ is negligible. But, in the present work, type-II seesaw mass term shows considerable contribution towards the said phenomenology. The exact reason although not known, this may be thought of as a result of emerging modular forms in the picture which changes the structure of the mass matrices within the model.
\begin{center}
\section{\label{lrsm7}Conclusion}
\end{center}
The experimental verification of neutrino oscillations paved the gateway for physics beyond the Standard Model. In the present work, LRSM has been realized with the help of modular $A_4$ symmetry, the advantage being, no extra particles called 'flavons'. The Yukawa couplings are represented as modular forms expressed as expansions of $q$. The values of the Yukawa couplings $(Y_1,Y_2,Y_3)$ are calculated using the relation $M_\nu^{II(diag)} = XM_\nu^{(diag)}$. The mass matrix for the type-II seesaw dominance has been determined using the multiplication rules for $A_4$ group stated in the Appendix. The Majorana mass matrix is found to be symmetric and under the considered basis, the charged lepton mass matrix is also diagonal. To summarize our work, some results are stated as under,
	\begin{itemize}
		\item In the present work we have considered the arbitrary constant that can be multiplied with the Yukawa term to be one. This is to reduce the number of free parameters in the model and to determine the data more accurately.
		\item The model has been realized using $A_4$ modular symmetry, the advantage being no use of extra particles or flavons. We can also use other symmetry groups depending upon the level of modular form used and as such, the number of modular forms will also change, as per the table \ref{table:3} in the manuscript. As we have used $A_4$ modular symmetry, there are three modular forms $(Y_1,Y_2,Y_3)$ and hence it became easier for us to construct the $3 \times 3$ mass matrices in terms of the aforementioned modular forms which are consistent with the neutrino data. 
		\item The Yukawa couplings determined for the cases when $X = 0.3, 0.5, 0.7$ has been found to satisfy the condition that its modulus is greater than unity.
		\item Effective masses corresponding to heavy right-handed neutrino contribution, and momentum dependent contributions of $0\nu\beta\beta$ are determined and their variations with the baryon asymmetry parameter $\eta_{B}$ has been studied. The results are found to satisfy the experimental constraints, for both normal and inverted hierarchies.
	\end{itemize}
	The effective masses for the $0\nu\beta\beta$ contributions are calculated and it has been found that the values for the effective mass corresponding to each contribution is well within the experimental bounds, which in fact makes us clearly state that the building of the model with modular symmetry is advantageous to that of flavor symmetries. In this model, no extra particles have been used and the calculations are carried out taking into picture the usual particle content of the model and hence conclusively it brings to our attention that different phenomenology in the Beyond Standard Model framework can be studied using modular symmetry that helps in keeping the model minimal.
	
\begin{center}
	\section*{Appendix A : Minimisation of the potential term.}
\end{center}

Let us consider the Higgs potential of our model that has quadratic and quartic coupling terms given by \cite{Luo:2008rs},
\begin{equation*}
	V_{\phi,\Delta_L,\Delta_R} = -\mu_{ij}^2 Tr[\phi_i^\dagger \phi_j] +\lambda_{ijkl} Tr[\phi_i^\dagger\phi_j]Tr[\phi_k^\dagger \phi_l]+ \lambda_{ijkl}^{'} Tr[\phi_i^\dagger\phi_j\phi_k^\dagger \phi_l] -\mu_{ij}^2 Tr[\Delta_L^\dagger \Delta_L + \Delta_R^\dagger \Delta_R] 
\end{equation*}
\begin{equation*}
	\rho_1 [(Tr[\Delta_L^\dagger \Delta_L])^2 + (Tr[\Delta_L^\dagger \Delta_L])^2] + \rho_2 (Tr[\Delta_L^\dagger \Delta_L \Delta_L^\dagger \Delta_L] + Tr[\Delta_R^\dagger \Delta_R \Delta_R^\dagger \Delta_R]) +\rho_3 (Tr[\Delta_L^\dagger \Delta_L] Tr[\Delta_R^\dagger \Delta_R]) + 
\end{equation*}
\begin{equation}
	\label{e:57}
	\alpha_{ij} Tr[\phi_i^\dagger \phi_j](Tr[\Delta_L^\dagger \Delta_L] + Tr[\Delta_R^\dagger \Delta_R]) + \beta_{ij}(Tr[\Delta_L^\dagger \Delta_L \phi_i \phi_j^\dagger] + (Tr[\Delta_R^\dagger \Delta_R \phi_i^\dagger \phi_j]) + \gamma_{ij}(Tr[\Delta_L^\dagger \phi_i \Delta_R \phi_j^\dagger] +h.c)
\end{equation}
where, i,j,k,l runs from 1 to 2 with $\phi_1 = \phi$ and $\phi_2 = \tilde{\phi}$.
As mentioned above after SSB, the scalar sector obtains VEV. So after the substitution of the respective VEVs and determining the traces, so after simplification the potential can be written as,
\begin{equation}
	\label{e:58}
	V = -\mu^2 (v_L^2 + v_R^2 ) + \frac{\rho}{4} (v_L^4 + v_R^4) + \frac{\rho'}{2} + \frac{\alpha}{2}(v_L^2 + v_R^2) k^2 + \gamma v_L v_R k^2
\end{equation}
where, we have used the approximation $k' << k$, and $\rho' = 2\rho_3$.
Our minimization conditions are, $\frac{\delta V}{\delta v_L} = \frac{\delta V}{\delta v_R} = \frac{\delta V}{\delta k} = \frac{\delta V}{\delta k'} = 0$

Therefore, we get,
\begin{equation}
	\frac{\delta V}{\delta v_L} = -2\mu^2 v_L + \rho v_L^3 + \rho' v_L k^2 + \gamma v_R k^2
\end{equation}
Here, it is evident that the Majorana mass of the left-handed neutrino $M_{LL}$ is dependent on the vev $v_L$ as already defined above.
Again, we have
\begin{equation}
	\frac{\delta V}{\delta v_R} = -2\mu^2 v_R + \rho v_R^3 + \rho' v_R k^2 + \gamma v_L k^2
\end{equation}
So, the right handed Majorana mass $M_{RR}$ is dependent on the vev $v_R$. Similarly, the calculations for the same can be carried out and it can be found out the Dirac mass term $M_D$ can be expressed in terms of the vev for the Higgs bidoublet as also defined previously.\\
Now, we are to determine a relation between the VEVs for the scalars and so after using the minimization conditions and simplifying the equations, we come to a relation given by,
\begin{equation}
	v_L v_R = \frac{\gamma}{\xi} k
\end{equation}
where, $\xi = \rho - \rho'$.\\ 
The neutrino mass for LRSM is given as a summation of the type-I and type-II term as already mentioned above. So, in the approximation that $k'<< k$, and if we consider that our Yukawa coupling $Y^l$ corresponding to the neutrino masses is $y_D$ and the coupling $\widetilde{Y^l}$ for the charged fermion masses is denoted by $y_L$, so considering $y_D k >> y_l k'$  we can write,
\begin{equation}
	M_\nu = \frac{k^2}{v_R} y_D f_R^{-1} y_D^T + f_L v_L
\end{equation}
Since, for due to left-right symmetry, we can consider $f_L = f_R =f$, so the above equation can be written as,
\begin{equation}
	M_\nu = \frac{k^2}{v_R} y_D f^{-1} y_D^T + f v_L
\end{equation}
So, from this equation we can come to a relation given by,
\begin{equation}
	M_\nu = (f\frac{\gamma}{\xi} + y_D f^{-1} y_D^T)\frac{k^2}{v_R}
\end{equation}
Here, we can consider two situations, namely
\begin{itemize}
	\item If $f(\frac{\gamma}{\xi}) << y_D f^{-1}y_D^T$, the light neutrino mass is given by the type-I term $M_D M_{RR}^{-1}M_D^T$. That is, here type-I is dominant and the light neutrino mass is from the suppression of heavy $\nu_R$.
	\item If $f(\frac{\gamma}{\xi}) >> y_D f^{-1}y_D^T$, the light neutrino mass is given by the type-II term $f v_L$. That is, in this case type-II mass term is dominant and the light neutrino mass is because of the tiny value of $\nu_L$.
\end{itemize}

\begin{center}
	\section*{Appendix B : Properties of $A_4$ discrete symmetry group.}
\end{center}

$A_4$ is a non-abelian discrete symmetry group which represents even permuatations of four objects. It has four irreducible representations, three out of which are singlets $(1,1',1'')$ and one triplet $3$ ($3_A$ represents the anti-symmetric part and $3_S$ the symmetric part). Products of the singlets and triplets are given by,
\begin{center}
	\begin{equation*}
		1 \otimes 1 = 1
	\end{equation*}
\end{center}
\begin{center}
	\begin{equation*}
		1' \otimes 1' = 1''
	\end{equation*}
\end{center}
\begin{center}
	\begin{equation*}
		1' \otimes 1'' = 1
	\end{equation*}
\end{center}
\begin{center}
	\begin{equation*}
		1'' \otimes 1'' = 1'
	\end{equation*}
\end{center}
\begin{center}
	\begin{equation*}
		3 \otimes 3 = 1 \oplus 1' \oplus 1'' \oplus 3_A \oplus 3_S
	\end{equation*}
\end{center}
If we have two triplets under $A_4$ say, $(a_1,a_2,a_3)$ and $(b_1,b_2,b_3)$ , then their multiplication rules are given by,
\begin{center}
	\begin{equation*}
		1 \approx a_1b_1 + a_2b_3 + a_3b_2
	\end{equation*}
	\begin{equation*}
		1' \approx a_3b_3 + a_1b_2 + a_2b_1
	\end{equation*}
	\begin{equation*}
		1'' \approx a_2b_2 + a_3b_1 + a_1b_3
	\end{equation*}
	\begin{equation*}
		3_S \approx \begin{pmatrix}
			2a_1b_1-a_2b_3-a_3b_2 \\
			2a_3b_3-a_1b_2-a_2b_1 \\
			2a_2b_2-a_1b_3-a_3b_1
		\end{pmatrix}
	\end{equation*}
	\begin{equation*}
		3_A \approx \begin{pmatrix}
			a_2b_3-a_3b_2 \\
			a_1b_2-a_2b_1 \\
			a_3b_1-a_1b_3
		\end{pmatrix}
	\end{equation*}
\end{center}

\begin{center}
	\section*{Appendix C : Modular Symmetry.}
\end{center}

Modular symmetry has gained much importance in aspects of model building \cite{King:2020qaj}, \cite{Novichkov:2019sqv}. This is because it can minimize the extra particle called 'flavons' while analyzing a model with respect to a particular symmetry group. An element $q$ of the modular group acts on a complex variable $\tau$ which belongs to the upper-half of the complex plane given as \cite{Novichkov:2019sqv} \cite{Feruglio:2017spp} 
\begin{equation}
	\label{E:23}
	q\tau = \frac{a\tau+b}{c\tau+d}
\end{equation}
where $a,b,c,d$ are integers and $ad-bc=1$, Im$\tau$$>$0.\\
The modular group is isomorphic to the projective special linear group PSL(2,Z) = SL(2,Z)/$Z_2$ where, SL(2,Z) is the special linear group of integer $2\times2$ matrices having determinant unity and $Z_2=({I,-I})$ is the centre, $I$ being the identity element. The modular group can be represented in terms of two generators $S$ and $T$ which satisfies $S^2=(ST)^3=I$. $S$ and $T$ satisfies the following matrix representations:
\begin{equation}
	\label{E:24}
	S = \begin{pmatrix}
		0 & 1\\
		-1 & 0
	\end{pmatrix}
\end{equation}

\begin{equation}
	\label{e:25}
	T = \begin{pmatrix}
		1 & 1\\
		0 & 1
	\end{pmatrix}
\end{equation}

corresponding to the transformations,
\begin{equation}
	\label{e:26}
	S : \tau \rightarrow -\frac{1}{\tau} ; T : \tau \rightarrow \tau + 1
\end{equation}
Finite modular groups (N $\leq$ 5) are isomorphic to non-abelian discrete groups, for example, $\Gamma(3) \approx A_4$, $\Gamma(2) \approx S_3$, $\Gamma(4) \approx S_4$. While using modular symmetry, the Yukawa couplings can be expressed in terms of modular forms, and the number of modular forms present depends upon the level and weight of the modular form. For a modular form of level N and weight 2k, the table below shows the number of modular forms associated within and the non-abelian discrete symmetry group to which it is isomorphic \cite{Feruglio:2017spp}.
\begin{table}[H]
	\begin{center}
		\begin{tabular}{|c|c|c|}
			\hline
			N & No. of modular forms & $\Gamma(N)$ \\
			\hline
			2 & k + 1 & $S_3$ \\
			\hline
			3 & 2k + 1 & $A_4$ \\
			\hline
			4 & 4k + 1 & $S_4$ \\
			\hline
			5 & 10k + 1 & $A_5$ \\
			\hline 
			6 & 12k &  \\
			\hline
			7 & 28k - 2 & \\
			\hline
		\end{tabular}
		\caption{\label{table:3}No. of modular forms corresponding to modular weight 2k.}
	\end{center}
\end{table}
In our work, we will be using modular form of level 3, that is, $\Gamma(3)$ which is isomorphic to $A_4$ discrete symmetry group. The weight of the modular form is taken to be 2, and hence it will have three modular forms $(Y_1,Y_2,Y_3)$ which can be expressed as expansions of q given by,
\begin{equation}
	\label{e:27}
	Y_1 = 1 + 12 q + 36 q^2 + 12 q^3 + 84 q^4 + 72 q^5 + 36 q^6 + 96 q^7 + 
	180 q^8 + 12 q^9 + 216 q^{10}
\end{equation}
\begin{equation}
	\label{e:28}
	Y_2 = -6 q^{1/3} (1 + 7 q + 8 q^2 + 18 q^3 + 14 q^4 + 31 q^5 + 20 q^6 + 
	36 q^7 + 31 q^8 + 56 q^9)
\end{equation}
\begin{equation}
	\label{e:29}
	Y_3 = -18 q^{2/3} (1 + 2 q + 5 q^2 + 4 q^3 + 8 q^4 + 6 q^5 + 14 q^6 + 
	8 q^7 + 14 q^8 + 10 q^9)
\end{equation}
where, $q = \exp(2\pi i \tau)$.\\

\section*{Acknowledgements}

Ankita Kakoti acknowledges Department of Science and Technology (DST), India (grant DST/INSPIRE Fellowship/2019/IF190900) for the financial assistantship.
\vspace{1cm}
\begin{center}
	\section*{References}
\end{center}

\bibliography{cite3}
\bibliographystyle{utphys}

\end{document}